\documentclass[aps,pre,amsmath]{revtex4}
\usepackage[a4paper]{geometry}
\usepackage{pstcol,color,pst-plot}
\usepackage[dvips]{graphicx}
\usepackage{verbatim}
\usepackage{amssymb} 
\usepackage{bm}
\usepackage{graphicx,color}
\usepackage{pstricks,pst-node}
\newcommand{\esc}{\!\cdot\!}
\newcommand{\hadr}{{\tt H-AdResS\;}}
\newcommand{\adr}{{\tt AdResS\;}}

\newcommand{\calf}{{\cal F}}
\newcommand{\ffec}{{F_{\mathrm{c}}}}
\newcommand{\lh}{{l}_{hyb}}

\usepackage{soul}
\begin{document}

\title{Statistical Mechanics of Hamiltonian Adaptive Resolution Simulations}
\author{P. Espa\~{n}ol$^{1}$}
\author{R. Delgado-Buscalioni$^2$}
\author{R. Everaers$^3$}
\author{R. Potestio$^4$}
\author{D. Donadio$^4$}
\author{K. Kremer$^4$}

\affiliation{$^1$Dept.   F\'{\i}sica Fundamental, Universidad  Nacional de
  Educaci\'on  a  Distancia, UNED,  Aptdo.   60141 E-28080,  Madrid,  Spain}
\affiliation{$^2$Dept.   F\'{\i}sica Te\'orica de la Materia Condensada, 
Universidad Aut\'onoma de Madrid and Institute for Condensed Matter Physics, IFIMAC. Campus de Cantoblanco, Madrid.}
\affiliation{$^3$Laboratoire de Physique et Centre Blaise Pascal, {\'E}cole Normale Sup\'erieure de Lyon, CNRS UMR5672, 46 all\'{e}e d'Italie, 69364 Lyon, France}
%\email{ralf.everaers@ens-lyon.fr}
\affiliation{$^4$ Max-Planck-Institut f\"ur Polymerforschung, Ackermannweg 10, 55128 Mainz, Germany}

%\date{27 March 2014}
%21 March 2014, 18 March 2014, 22 may 2013,4th April 2013, 19 July 201210 July 2012, 23 june 2012, 11 june 2012, 23 May 2012, 28 April 2012
%21 march 2013
\begin{abstract}
  The Adaptive Resolution Scheme (\adr) is a  hybrid scheme  that allows
  to  treat  a molecular system with different levels of resolution depending on the location
  of the  molecules.  The construction  of a Hamiltonian based  on the
  this idea (\hadr) allows one to formulate the usual tools of ensembles and
  statistical mechanics. We present  a number of exact and approximate
  results  that provide  a statistical  mechanics foundation  for this
  simulation   method.  We  also   present  simulation   results  that
  illustrate the theory.
\end{abstract}
\maketitle
\section{Introduction}

Biological and soft matter  systems are characterized by the existence
of  processes  with many  different  length  and  time scales.   These
processes are usually coupled, making their theoretical, experimental,
and computer simulation description  a daunting task.  The functioning
of a protein, for example, involves chemical processes at active sites
as well  as the  overall dynamics of  the protein and  its environment
\cite{Neri2005}.  Crack  propagation is  another example in  which the
atomic  processes  occuring at  the  crack  tip  affect crucially  the
overall   elastic    behaviour   of   the    sample,   and   vice-versa
\cite{Csanyi2004c}.   From a  computational point  of view,  the brute
force approach  of treating the  system with full molecular  detail is
not possible, and one needs to deal with simplified, or coarse-grained
versions of  the system \cite{Voth2009}.  By  definition, in any  coarse-grained model
some atomic/molecular  detail is lost.   In some fortunate  cases, the
need for atomistic detail is confined  in small regions of space as in
the examples  above, and there is  hope that a  hybrid scheme coupling
all atom (AA) with coarse-grained (CG) descriptions may be a successful
approach. The  coupling of different models describing  the system at
different resolutions  is an active field  of research \cite{Park2009,Wang2013}
and in our opinion it will become useful in a broad range of calculations,
beyond the multi-scale community \cite{Wang2013}.

We have recently developed an Hamiltonian Adaptive Resolution Scheme (\hadr)
\cite{Potestio:PRL:110,Potestio:PRL:111}.     Other    proposals   for
Hamiltonian hybrid (AA/CG) schemes have been presented \cite{Park2009}
which are  technically challenging as compared with  \hadr. As opposed
to previous  versions of AdResS, where  a \textit{force} interpolation
principle was  the crucial  element, in \hadr  \textit{potentials} are
interpolated.  The proposed Hamiltonian  in \hadr includes a switching
field  that  allows  for  a  swift  interpolation  between  the  truly
microscopic  Hamiltonian and  a CG  version  of it.   When a  molecule
crosses the interface  between the AA and CG  regions, its interaction
with other molecules changes accordingly.  Usually the CG potential of interaction 
used in the CG region is only an approximate version  of the actual potential of mean force.
The discrepancies between  the CG potential and the  potential of mean
force are taken  into account in the \hadr  Hamiltonian through a {\em
  free energy compensation term} \cite{Potestio:PRL:110,Potestio:PRL:111}.

The idea of interpolating AA  an CG potentials through a hybrid region
is      not     new      and     was      introduced      in     Refs.
\cite{Ensing2007},\cite{Nielsen2010}   under  the  name   of  adaptive
Multiscale  Molecular Dynamics (MMD)\cite{Nielsen2010}.   However, the
detailed form of the interpolation  is slightly different in \hadr and
leads to the  existence of a well-defined Hamiltonian  that allows for
the natural  use of the  principles of Statistical Mechanics.   In the
original    version    of    MMD,    energy    was    not    conserved
\cite{DelleSite2007},\cite{Praprotnik2011a}   and   thermostats   were
required     \cite{Ensing2007},\cite{Nielsen2011}.    In  the original
thermostatted \adr \cite{Matej_JCP07} and also in more recent versions \cite{Wang2013},
the  mass  in  the   atomistic  domain  fluctuates  according  to  the
Grand-Canonical  ensemble; at  least up  to the  second moment  of the
probability    density     function, as    it     has    numerically
\cite{Delgado-Buscalioni2008},  and theoretically shown
\cite{Wang2013}. 
Density  fluctuations are  determined  by the  fluid
compressibility, specified by the  integral of the radial distribution
function, and  by finely  tunning the CG  potential one can  match the
compressilibities  of  the  CG   and  AA  domains.   Having  the  same
compressibility does not however ensure the same pressure equation of state
and to ensure a constant density profile over the CG and AA domains, a
recent work \cite{FritschPRL108} proposes the imposition of a ``correction force  field'', which
is iteratively evaluated according to the idea of imposing pressure  balance 
(and thus involving compressibilities).  The existence of a Hamiltonian permits 
us to derive a fundamental relation between  the force density  and the
density  gradient,   which  turns  out   to  be  independent   on  the
compressibility.  This relation explains the basis of the ``correction
force   field''    used   to   control   the    density   profile, not only
in Ref.  \cite{FritschPRL108}   but also in   many  other  algorithms using domain
decomposition (see  e.g. Refs. \cite{Issa2014,kotsalis07}).

The Hamiltonian
system of \hadr also allows to extend the working ensemble also to the
$E,N,V$  microcanonical  ensemble,  with  no need  of  any  extraneous
thermostats. However,  the benefits of a  Hamiltonian description will
prove   to   be   substantially   broader,   as   already   shown   in
Ref. \cite{Potestio:PRL:111}.

In the present  paper, we derive the Statistical  Mechanics basis for
the \hadr method. Several exact results concerning the local equations
of state for the pressure and temperature allow for the formulation of the
free energy compensation term in  an iterative way.  We also show that
under a local equilibrium approximation, valid when the hybrid region is wide, 
the iterative procedure can be simplified leading  to an approximate  but very efficient way  for the
calculation of  the free energy compensation term  in the Hamiltonian.
We have analyzed the effect  of the width of the transition layer
where molecules gradually change  their resolution. A relevant outcome
is that the \hadr {\em  total} free energy compensation is independent
on  the layer,  even for  widths of  the same  order of  the molecular
diameter.  Another very important  observation is that the \hadr total
free energy correction is equal, within error bars, to the free energy
difference   between  both   fluids  (atomistic   and  coarse-grained)
evaluated from Kirkwood thermodynamic integration \cite{Kirkwood1935}.
Although more research is required in this direction, this would allow
\hadr to  be used  as a  flexible tool for  estimation of  free energy
differences in different scenarios.

In what follows, we first present the \hadr Hamiltonian formulation in
Sec.   \ref{sec:ham}.   The free  energy  corresponding  to the  \hadr
Hamiltonian is introduced in Sec. \ref{sec:free}. In Sec. \ref{sec:PT}
we derive expressions for the temperature and the pressure tensor fields.  In
Sec. \ref{sec:trans}  we demonstrate  that the condition  of constant
pressure  over  the  \hadr  simulation  stems from  the  condition  of
translational  invariance  of  the  free energy.   The  force  balance
equation derived  in Sec.  \ref{sec:itera} permits  to rationalize the
different  types  of \hadr  compensation  terms,  for either  constant
pressure  or density  fields.   Section \ref{sec:lea}  shows that  under local
equilibrium (LE) conditions, the FEC  is just the Kirkwood free energy
difference  \cite{Kirkwood1935},  thus  justifying  the  non-iterative
route         used        in         our         previous        works
\cite{Potestio:PRL:110,Potestio:PRL:111}.   Finally,  the  theoretical
framework    is     validated    through    simulations     in    Sec.
\ref{sec:simulations} where  we also provide  relaxational schemes for
the  iterative route  to the  FEC.  We  also study  the effect  of the
transition   layer  width   and  the   deviation  from   the  Kirkwood
approximation to  the FEC.   Conclusions and some  future perspectives
are given in Sec. \ref{sec:conclu}.

\section{The AdResS Hamiltonian }
\label{sec:ham}
Consider  a classic molecular system  composed of  $N$ constituent  atoms. The
microscopic  state of  the system  is described  by the  positions and
momenta  of the  atoms, denoted  generically by  $r,p$. The  system is
coarse-grained by considering the centers  of mass (CoM) of $M$ groups
of atoms that  are bound together and that are  termed {\em blobs}.  A
blob may  be, for  example, a single  molecule or  a part of  a bigger
molecule.  The position of the  $\mu$-th blob CoM is $\hat{\bf R}_\mu$
which is defined as the following phase function
\begin{align}
  \hat{\bf R}_\mu(r)&=\sum_i^N\delta_\mu(i){\bf r}_i\frac{m_i}{M_\mu}=\sum^{N_\mu}_{i_\mu}{\bf r}_{i_\mu}\frac{m_{i_\mu}}{M_\mu}
\nonumber\\
M_\mu&=\sum_i^N\delta_\mu(i)m_i
\label{1}
\end{align}
where the indicator  symbol $\delta_\mu(i)$ takes the value  1 if atom
$i$ is  in blob $\mu$ and  zero otherwise.  The  last definition makes
use of  the notation  $i_\mu$ that corresponds  to the $i$-th  atom of
blob $\mu$  and ${N_\mu}$ is the  number of atoms of  blob $\mu$.  The
microscopic Hamiltonian governing the dynamics of the atoms is
\begin{align}
  H^1(r,p) &=\sum_{i}^{N}\frac{{\bf p}_{i}^2}{2m_{i}}+\sum_\mu^MV^{\rm intra}_\mu(r)+V^1(r)
\label{H1}
\end{align}
where  the total  potential  energy  of interaction  of  the atoms  is
decomposed into  the potential of  interaction between atoms  within a
blob $V^{\rm  intra}_\mu(r)$ and the potential  of interaction between
atoms  of  different  blobs  $V^1$.   This  potential  energy  can  be
decomposed as  $V^1=\sum_\mu^MV^1_\mu$ where the  terms $V^1_{\mu}(r)$
are  the potential  energy of  interaction of  the atoms  of different
blobs  where one of the atoms of the  pair is in blob
$\mu$. Explicitly
\begin{align}
  V^{1}_{\mu}(r)&=\frac{1}{2}\sum_{ij}^N \delta_\mu(i)\phi^{\rm inter}(r_{ij})
\end{align}
where $\phi^{\rm inter}$ is the  pair potential between atoms $i,j$ of
different blobs. It  is understood that $\phi^{\rm  inter}(r_{ij})$ is zero
if atoms $i,j$ belong to  the same blob. Note that any Hamiltonian
that differs  from the one in  Eq. (\ref{H1}) by a  constant term will
produce exactly the  same dynamics.  The usual convention  is to chose
the zero of potential energy in such a way that when the particles are
very far apart and,  therefore, non-interacting, the potential energy
is zero.   This fixes the origin  of the energy scale.  We will assume
that  the above  Hamiltonian has,  through  a mixing  property of  its
Hamiltonian flow, a well defined  equilibrium ensemble. For this to be
true it  is necessary  that the  {\em finite} system  of $N$  atoms were
confined, either by a time-independent external field (not included in
(\ref{H1})) or through periodic boundary conditions.

The  central  idea  of  \hadr   is  to  introduce  a  switching  field
$\lambda({\bf r})$ that takes the value 1 in the region of space where
the system is described in full  all atomic (AA) detail, and the value
0  in  the  region  of  space  where the  system  is  described  in  a
coarse-grained  (CG) way.  In  the transition  region between  the two
zones the switching field changes  monotonously from 0 to 1. The field
$\lambda({\bf  r})$ gives  the degree  of detail  of  the description.
Instead of the microscopic Hamiltonian (\ref{H1}), the dynamics of the
atoms is modified with the following \hadr Hamiltonian,
\begin{align}
  H_{[\lambda]}(r,p)&=\sum_i^N \frac{{\bf p}_i^2}{2m_i}+V_{[\lambda]}(r)
\nonumber\\
V_{[\lambda]}(r)&=\sum_\mu^M V^{\rm intra}_\mu(r)+
\sum_{\mu}^M\lambda(\hat{\bf R}_\mu)V^{1}_{\mu}(r)
\nonumber\\
&+\sum_{\mu}^M(1-\lambda(\hat{\bf R}_\mu))V^0_{\mu}(R)
+\sum_\mu^M\calf(\lambda(\hat{\bf R}_\mu))
\label{HA}
\end{align}
The  potential  $V^{0}_\mu(R)$ is  assumed  to  depend  on the  atomic
coordinates  $r$ {\em  only} through  the position  of the  centers of
mass,        denoted        collectively       as        $R=\{\hat{\bf
  R}_\mu(r),\mu=1,\cdots,M\}$.   Although the present formalism is general and allows
for multi-body CG potentials, in most practical cases a  pair-wise  form  will  be assumed, this is
\begin{align}
  V^{0}_\mu(R)&=\frac{1}{2}\sum_{\nu}^MV^{0}(\hat{\bf R}_\mu-\hat{\bf R}_\nu)
\label{pw}
\end{align}
The potential $V^{0}_\mu(R)$ describes the interaction between blobs in
a coarse-grained way.
The term  $\sum_\mu^M\calf(\lambda(\hat{\bf R}_\mu))$  in the
Hamiltonian is referred to as {\em the  free energy compensation
  term}. Its effect is very much  like an external field acting on the
blobs. We require that $\calf(1)=0$.

The   rationale  for   postulating  the   above  Hamiltonian   is  the
following. When $\lambda({\bf r})  =1$ the above Hamiltonian coincides
with  the microscopic Hamiltonian  (\ref{H1}), this  is $H_{[1]}=H^1$.
On the other hand, when $\lambda({\bf r})=0$ the Hamiltonian becomes
\begin{align}
  H_{[0]}(r,p)&=\sum_i^N\frac{{\bf p}_i^2}{2m_i}+\sum_\mu^M V^{\rm intra}_\mu(r)
\nonumber\\
&+
\sum_{\mu}^MV^{0}_{\mu}(R)
+\sum_\mu^M\calf(0)
\label{HAl=0}
\end{align}
where,  apart from the  constant term  $\sum_\mu^M\calf(0)$,
the potential of interaction between atoms of different blobs is given
by the CG  interaction.  Therefore, the idea is  that with a spatially
varying $\lambda({\bf  r})$ the blobs change its  interaction from its
real microscopic  interaction $V^{1}(r)$  to a CG  interaction through
its  centers of  mass $V^{0}(R)$.  In  fact, the  equations of  motion
produced by  the Hamiltonian (\ref{HA}) are (assume  that particle $i$
belongs to blob $\mu$)
\begin{align}
\dot{\bf r}_i &=\frac{{\bf p}_i}{m_i}
\nonumber\\
\dot{\bf p}_i &= -\frac{\partial V_\mu^{\rm intra}}{\partial {\bf r}_i}
-
\sum^M_{\nu}\lambda(\hat{\bf R}_\nu)\frac{\partial V_\nu^{1}}{\partial {\bf r}_i}
-\sum^M_{\nu}(1-\lambda(\hat{\bf R}_\nu))\frac{\partial V^{0}_{\nu}}{\partial {\bf r}_i}
\nonumber\\
&-\nabla \lambda({\bf R}_\mu)\frac{m_i}{m_\mu}\left(V^{1}_{\mu}-V^{0}_{\mu}+
\calf'(\lambda(\hat{\bf R}_\mu))
\right),
\label{eqmot1}
\end{align}
where the  prime ($\calf' = d\calf/d\lambda$)  denotes derivative with
respect to $\lambda$.  When  $\lambda=1$ Eq. (\ref{eqmot1}) correspond to
the fully resolved microscopic dynamics, this is
\begin{align}
\dot{\bf r}_i &=\frac{{\bf p}_i}{m_i}
\nonumber\\
\dot{\bf p}_i &= -\frac{\partial V_\mu^{\rm intra}}{\partial {\bf r}_i}
-\sum^M_{\nu}\frac{\partial V_\nu^{1}}{\partial {\bf r}_i}
\end{align}
When $\lambda=0$  Eq. (\ref{eqmot1}) become  
\begin{align}
\dot{\bf r}_i &=\frac{{\bf p}_i}{m_i}
\nonumber\\
\dot{\bf p}_i &= -\frac{\partial V_\mu^{\rm intra}}{\partial {\bf r}_i}
-\sum^M_{\nu}\frac{\partial V^{0}_{\nu}}{\partial {\bf r}_i}
\end{align}
that  describes  the  motion  of  the  atoms  as  given  in  terms  of
microscopic  forces  due  to  the  atoms  of  the  same  blob  and  CG
interactions between  the centers of mass  of the blobs.  In this way,
{\em in the CG region, the Hamiltonian of \hadr moves the atoms with CG
  interactions.}

In  the transition  region when  $0<\lambda<1$ the  atoms move  with a
combination  of the microscopic  and CG  potentials and,  in addition,
feel the  presence of an  ``external field'', represented in  the last
term of the momentum equation (\ref{eqmot1}), which is proportional to
the  gradient of $\lambda$.   The contribution  $\calf'(\lambda)$ that
appears in Eq.  (\ref{eqmot1}) has the mission to
make  this  ``external field''  effect  as  small  as possible,  in  a
statistical sense.  We  will give in the next  section a thermodynamic
interpretation   to   the   $\calf(\lambda)$   contribution   in   the
Hamiltonian.   A molecular  dynamics simulation  with  the Hamiltonian
(\ref{HA}) can  be coded  in a way  that the simulation  proceeds much
faster  than  the  one  given  by  the  full  microscopic  Hamiltonian
(\ref{H1}).  Indeed, in  the CG region the forces on  the atoms need a
search  only of  the neighbouring  {\em  blobs} whose  number is  much
smaller  than  the  number   of  atoms  required  in  the  microscopic
evaluation and indeed in the CG domain,
the number of force evaluations is drastically reduced.

Note that the  way in which the AA and  CG potentials are interpolated
in the  Hamiltonian (\ref{HA}) is different from  the interpolation in
the  MMD  method  \cite{Ensing2007,Nielsen2011}  where in  the  latter
method the switching  function depends on the position  of the centers
of mass of two blobs instead of just one blob in \hadr.

\section{ The free energy}
\label{sec:free}
The thermodynamic free energy  corresponding to the AdResS Hamiltonian
(\ref{HA}) is given by the usual statistical mechanics formula
\begin{align}
  F_{[\lambda]}&=-k_BT\ln\int{d^{3N}rd^{3N}p}\exp\left\{-\beta
H_{[\lambda]}(r,p)\right\}
\nonumber\\
&=-k_BT\ln\int \frac{d^{3N}r}{\Lambda^{3N}}\exp\left\{-\beta
V_{[\lambda]}(r)\right\}
\label{Flam}
\end{align}
and it is a functional  of the switching field $\lambda({\bf r})$.  In
this expression  the momentum integrals  of the kinetic energy  in the
Hamiltonian   have  been   performed   giving  rise   to  the   factor
$\Lambda^{3N}$
\begin{align}
  \Lambda^{3N}\equiv\prod_{\mu}^M\prod_{i_\mu}^{N_\mu}\Lambda_{i_\mu}^{3}
\end{align}
where the thermal wavelength of atom $i_\mu$ is defined as
\begin{eqnarray}
  \Lambda_{i_\mu}=\left(\frac{h^2}{2\pi k_BT m_{i_\mu}}\right)^{1/2}
\end{eqnarray}
The macroscopic thermodynamic free energy can be expressed in
terms of a potential of mean force by introducing the identity in
the form
\begin{align}
  1=\int  d^{3M}R\prod_\mu^M\delta({\bf  R}_\mu-\hat{\bf
  R}_\mu(r) )
\label{1delta}
\end{align}
Recall that $\hat{\bf R}_\mu(r) $ is a phase function that depends on the positions of the
atoms of blob $\mu$, i.e. Eq. (\ref{1}). 
\begin{widetext}
By inserting (\ref{1delta}) inside  the free energy (\ref{Flam}) leads to 
\begin{align}
  F_{[\lambda]}
% &=-k_BT\ln\int  d^{3M}R\exp\left\{-\beta\left[
% \sum_{\mu}^M(1-\lambda({\bf R}_\mu))V^{0}_{\mu}(R)
% +\sum_\mu^M\calf(\lambda({\bf R}_\mu))\right]\right\}
% \nonumber\\
% &\times\int\frac{d^{3N}r}{\Lambda^{3N}}\exp\left\{-\beta
% \left[V^{\rm intra}(r)+
% \sum_{\mu}^M\lambda({\bf R}_\mu)V^{1}_{\mu}(r)
% \right]\right\}\prod_\mu^M\delta({\bf R}_\mu-\hat{\bf R}_\mu)
% \nonumber\\
&=-k_BT\ln\int \frac{d^{3M}R}{\Lambda_0^{3M}}\exp\left\{-\beta\left[
\sum_{\mu}^M(1-\lambda({\bf R}_\mu))V^{0}_{\mu}(R)+V^{\rm mf}_{[\lambda]}(R)
+\sum_\mu^M\calf(\lambda({\bf R}_\mu))\right]\right\}
\label{Fm1}
\end{align}
where the potential  of mean force is defined as
% \begin{align}
% \exp\left\{-\beta V^{\rm mf}_{[\lambda]}(R)\right\}&\equiv\int\frac{d^{3N}r}{\Lambda^{3N}}\exp\left\{-\beta
% \left[V^{\rm intra}(r)+
% \sum_{\mu}^M\lambda({\bf R}_\mu)V^{1}_{\mu}(r)
% \right]\right\}\Lambda_0^{3M}\prod_\mu^M\delta({\bf R}_\mu-\hat{\bf R}_\mu)
% \end{align}
\begin{align}
V^{\rm mf}_{[\lambda]}(R)&\equiv-k_BT\ln \int\frac{d^{3N}r}{\Lambda^{3N}}\exp\left\{-\beta
\left[V^{\rm intra}(r)+
\sum_{\mu}^M\lambda({\bf R}_\mu)V^{1}_{\mu}(r)
\right]\right\} \Lambda_0^{3M} \prod_\mu^M\delta({\bf R}_\mu-\hat{\bf R}_\mu(r))
\label{Vmf}
\end{align}
$\Lambda_0$ is an  {\em arbitrary} length scale that renders the argument of
the logarithms in Eqs. (\ref{Fm1}) and (\ref{Vmf})  dimensionless.   

The  potential  of mean  force  (\ref{Vmf})  is  a functional  of  the
switching field  $\lambda$.  When $\lambda({\bf  r})=1$, the effective
potential $V^{\rm  mf}_{[1]}(R)$ coincides with the  potential of mean
force of the fully microscopic Hamiltonian $H_{[1]}(r,p)$, this is
\begin{align}
V^{\rm mf}_{[1]}(R)&\equiv-k_BT\ln \int\frac{d^{3N}r}{\Lambda^{3N}}\exp\left\{-\beta
\left[V^{\rm intra}(r)+
\sum_{\mu}^MV^{1}_{\mu}(r)
\right]\right\}\Lambda_0^{3M}\prod_\mu^M\delta({\bf R}_\mu-\hat{\bf R}_\mu)
\label{Vmf1}
\end{align}
On the other hand, when $\lambda({\bf r})=0$, we have
% \begin{align}
%   \exp\left\{-\beta        V^{\rm mf}_{[0]}(R)\right\}
%   &\equiv\int\frac{d^{3N}r}{\Lambda^{3N}}   \exp\left\{-\beta   V^{\rm
%       intra}(r)\right\}\Lambda_0^{3M}\prod_\mu^M\delta({\bf
%     R}_\mu-\hat{\bf R}_\mu)
%   \nonumber\\
%   &=\prod_\mu^M\int\frac{d^{3N_\mu}r}{\Lambda^{3N}}\exp\left\{-\beta
%     V_\mu^{\rm            intra}(r_\mu)\right\}\Lambda_\mu^3\delta({\bf
%     R}_\mu-\hat{\bf   R}_\mu)  =   \exp\left\{-\beta   \sum_\mu^M  F_\mu^{\rm
%     intra}\right\}
% \label{veff0}
% \end{align}  
\begin{align}
V^{\rm mf}_{[0]}(R)
  &\equiv-k_BT\ln \int\frac{d^{3N}r}{\Lambda^{3N}}   \exp\left\{-\beta   V^{\rm
      intra}(r)\right\}\Lambda_0^{3M}\prod_\mu^M\delta({\bf
    R}_\mu-\hat{\bf R}_\mu(r))
  \nonumber\\
  &=-k_BT\ln \prod_\mu^M\int\frac{d^{3N_\mu}r}{\Lambda^{3N}}\exp\left\{-\beta
    V_\mu^{\rm            intra}(r_\mu)\right\} \Lambda_0^3\delta({\bf
    R}_\mu-\hat{\bf   R}_\mu)  =   \sum_\mu^M  F_\mu^{\rm    intra}
\label{veff0}
\end{align}
where we  have introduced  the actual  thermodynamic free
energy  $F^{\rm  intra}_\mu$ that  a  blob  would  have should  it  be
isolated from the rest of blobs, this is
\begin{align}
 \exp\left\{-\beta F_\mu^{\rm intra}\right\}
&\equiv  \int\frac{d^{3N_\mu}r}{\Lambda^{3N_\mu}}
\exp\left\{-\beta V_\mu^{\rm     intra}(r_\mu)\right\}\Lambda_0^3
\delta({\bf R}_\mu-\hat{\bf R}_\mu)
\label{Fmu}\end{align}
\end{widetext}
Note  that, in spite  of the  appearance of  the Dirac  delta function in Eq. (\ref{Fmu})
depending on ${\bf R}_\mu$, this internal blob free energy $F_\mu^{\rm
  intra}$  is  independent  of   ${\bf  R}_\mu$  due  to  translational
invariance.   Therefore, we  may integrate  both sides  of (\ref{Fmu})
with respect to ${\bf R}_\mu$ leading to
\begin{align}
 \exp\left\{-\beta F_\mu^{\rm intra}\right\}& =\frac{\Lambda_0^3}{V}\int\frac{d^{3N_\mu}r}{\Lambda^{3N_\mu}}\exp\left\{-\beta
    V_\mu^{\rm     intra}(r_\mu)\right\}
\end{align}
where  $V$  is  the  total  volume  of  the  system.  

Therefore,  in  the  two  limits $\lambda({\bf  r})=1$,  $\lambda({\bf
  r})=0$, the free energy (\ref{Flam}) becomes
\begin{align}
  F_{[1]}
&=-k_BT\ln \int\frac{ d^{3M}R}{\Lambda_0^{3M}}\exp\left\{-\beta
V_{[1]}^{\rm mf}(R)\right\}
\nonumber\\
  F_{[0]}
&=-k_BT\ln\int \frac{ d^{3M}R}{\Lambda_0^{3M}}\exp\left\{-\beta
\sum_{\mu}^M\left[V^{0}_{\mu}(R)
+F^{\rm intra}_\mu\right]\right\}
\nonumber\\
&+M \calf(0)
\label{Fm4}
\end{align}
The requirement  of thermodynamic  consistency between both  levels of
resolution  enforces  that the  thermodynamic  free  energy should  be
exactly the same in both limits, that is,
\begin{align}
    F_{[0]}&=  F_{[1]}
\label{F0F1}
\end{align}
This thermodynamic consistency requirement  gives light to the meaning
of the free energy compensating  term $\calf (\lambda)$. In the spirit
of changing the resolution, we expect that $V_0(R)$ in Eq.  (\ref{HA})
is given by the potential of mean force of the microscopic Hamiltonian
(\ref{H1}). This potential of mean  force can be measured in different
ways,   from   Boltzmann  inversion  \cite{Faller2004}   to  relative   entropy
\cite{Shell2008a} methods.  These methods allow one to obtain $V_0(R)$
{\em up  to an arbitrary  constant}.  Indeed $V_0(R)$ is  a mesoscopic
free  energy for  which only  relative  values may  be computed.  This
constant is usually  fixed by requiring that $V_0(R)$  vanishes as the
centers   of    mass   become   apart,    i.e.    $|{\bf   R}_\mu-{\bf
  R}_\nu|\to\infty$.  On  the other hand, the potential  of mean force
$V_{[1]}^{\rm   mf}(R)$  of   the  microscopic   Hamiltonian  contains
information of not only the  interactions between blobs but also about
the  internal free energy  of the  molecules.  One  way in  which this
clearly manifests is when the blobs in which we have grouped the atoms
correspond to full molecules.  In that  case it makes sense to look at
the low density regime in which the molecules are very far from each
other.  In this  limit, we obtain from Eq.  (\ref{Vmf1}) that when the
centers of mass  are separated beyond the range  of interaction of the
potentials,  then  we  may   neglect  the  term  $V_\mu^1(r)$  in  Eq.
(\ref{Vmf1}),  leading  to  $V^{\rm  mf}_{[1]}(R)=\sum_\mu  F_\mu^{\rm
  intra}$.  As a  result, the potential of
mean  force $V_{[1]}^{\rm  mf}(R)$  does not  vanish  as the  distance
between particles  goes to  infinity, as opposed  to $V_0(R)$.   If we
momentarily assume that the  many-body potential of mean force $V^{\rm
  mf}_{[1]}(R)$ could  be very well approximated by  a pair-wise form,
we would choose the  pair-wise potential $V^{0}(R)$ as $V^{0}(R)=V^{\rm
  mf}_{[1]}(R)-\sum_\mu  F_\mu^{\rm  intra}$  (vanishing  as  the  CoM
separate).  In that  situation, the consistency (\ref{F0F1}) would
imply  $\calf(0)=0$.  It  is clear,  therefore, that  the contribution
$\calf(0)$   has  the   effect  of   ``curing'',  at   the   level  of
thermodynamics, the errors due to  the use of an approximate pair-wise
potential  $V_0(R)$ for the  actual many-body  potential of  mean force
$V_{[1]}^{\rm mf}(R)$.

The  free energy  (\ref{Flam})  is  a functional  of  the switching  field
$\lambda({\bf r})$.  For future  reference, we compute  explicitly the
functional derivative of the free energy with respect to $\lambda({\bf
  r})$, this is
\begin{align}
\frac{\delta F_{[\lambda]}}{\delta\lambda({\bf r})}
&=\left\langle \frac{\delta H_{[\lambda]}}
{\delta \lambda({\bf r})}\right\rangle^{[\lambda]}
\label{dhdla}
\end{align}
In   this   expression,   $\langle\cdots\rangle^{[\lambda]}$  is   a
canonical   average  with   the  AdResS   Hamiltonian  $H_{[\lambda]}$
in Eq. (\ref{HA}).  By using
\begin{align}
 \frac{\delta H_{[\lambda]}}{{\delta \lambda({\bf r})}}&
=\left( u^1_{{\bf r}}-u^0_{{\bf r}}
+\calf'(\lambda({\bf r}))n_{{\bf r}}\,\right)
\label{dhdl}
\end{align}
where have  defined the  potential energy
densities  $ {u}^0_{\bf  r}, {u}^1_{\bf  r}$  and the  center of  mass
density ${n}_{\bf r}$ as
\begin{align}
  {u}^1_{\bf r}&\equiv\sum^M_\mu V^1_\mu\delta(\hat{\bf R}_\mu-{\bf r})
\nonumber\\
  {u}^0_{\bf r}&\equiv\sum^M_\mu V^{0}_\mu\delta(\hat{\bf R}_\mu-{\bf r})
\nonumber\\
{n}_{\bf r}&\equiv\sum^M_\mu\delta(\hat{\bf R}_\mu-{\bf r})
\label{defu1u0}
\end{align}
we finally obtain the explicit expression for the functional derivative of the free energy of \hadr
\begin{align}
\frac{\delta F_{[\lambda]}}{\delta\lambda({\bf r})}=\left\langle {u}^1_{\bf r}-{u}^0_{\bf r}\right\rangle^{[\lambda]}+
  \calf'(\lambda({\bf r}))  \left\langle {n}_{\bf r}\right\rangle^{[\lambda]}
\label{dfdl=01}
\end{align}
This expression will be used below.

\section{The temperature and pressure fields}
\label{sec:PT}
In the previous  section, we have presented a  consistency argument in
Eq.  (\ref{F0F1}) based  on  the global  thermodynamics  of the  \hadr
system.   In this  section we  formulate the  local  thermodynamics of
\hadr in terms  of the equations of state for  the temperature and the
pressure. In  order to achieve this,  it is convenient to  look at the
{\em  molecular} momentum  density field  because its  time derivative
will  give   information  about  mechanical   equilibrium  and,  hence,
pressure. The molecular momentum density field is defined as
\begin{align}
  \hat{\bf g}_{\bf r}(z) &\equiv\sum_\mu^M \hat{\bf P}_\mu\delta(\hat{\bf R}_\mu-{\bf r})
\label{hatg}
\end{align}
where the momentum $\hat{\bf P}_\mu$ of blob $\mu$ is given by
\begin{align}
  \hat{\bf P}_\mu(r)&=\sum_i^N\delta_\mu(i){\bf p}_i
\end{align}
The  time derivative  of the phase function (\ref{hatg}) is
obtained   by  applying   the  Liouville   operator  onto   this
function, providing
\begin{align}
  iL\hat{\bf g}_{\bf r}&=\hat{\bf f}_{\bf r} -\nabla\hat{\bf K}_{\bf r}
\label{lrlg}
\end{align}
where the {\em kinetic part of the stress tensor} is defined as
\begin{align}
  \hat{\bf K}_{\bf r}&\equiv\sum_\mu^M \hat{\bf P}_\mu\hat{\bf V}_\mu\delta(\hat{\bf R}_\mu-{\bf r}).
\label{kinst}
\end{align}
The velocity is $\hat{\bf V}_\mu=\hat{\bf P}_\mu/M_\mu$, and  the {\em force density} is defined as 
\begin{align}
  \hat{\bf f}_{\bf r}&\equiv \sum_\mu^M \hat{\bf F}_\mu\delta(\hat{\bf R}_\mu-{\bf r})
\label{fr}
\end{align}
Here, $\hat{\bf F}_\mu$ is the force on molecule $\mu$ which is given by
\begin{align}
\hat{\bf F}_\mu&\equiv-\sum_i\delta_\mu(i) \frac{\partial H_{[\lambda]}}{\partial {\bf r}_i}.
\label{fmu}
\end{align}
In the Appendix \ref{Ap:force}, it is  shown that the force ${\bf  F}_\mu$ on molecule $\mu$
introduced in Eq. (\ref{fmu}) has the following form
\begin{align}
\hat{\bf F}_\mu
&=\sum_{\nu}  \hat{\bf G}_{\mu\nu}
\nonumber\\
&-\nabla\lambda(\hat{\bf R}_\mu)(V^{1}_{\mu}(r)-V^{0}_{\mu}(R)+\calf'(\lambda_\mu(R)))
\end{align}
where we have introduced the pair force 
\begin{align}
  \hat{\bf G}_{\mu\nu}&\equiv
\left[\frac{\lambda(\hat{\bf R}_\mu)+\lambda(\hat{\bf R}_\nu)}{2}\right]
{\bf F}^{1}_{\mu\nu}(R_{\mu\nu})
\nonumber\\
&+\left[1-\frac{\lambda(\hat{\bf R}_\mu)+\lambda(\hat{\bf R}_\nu)}{2}\right]
{\bf F}^{0}_{\mu\nu}(R_{\mu\nu})
\end{align}
This  force  satisfies Newton's  Third  Law  $ \hat{\bf  G}_{\mu\nu}=-
\hat{\bf   G}_{\nu\mu}$.     The   forces   ${\bf   F}^1_{\mu\nu},{\bf
  F}^0_{\mu\nu}$  introduced  in   Appendix  \ref{Ap:force}  are  the
original microscopic and CG forces between blobs, respectively. We may
compute now the force density $\hat{\bf f}_{\bf r}$ in Eq.  (\ref{fr})
and obtain
\begin{align}
\sum_\mu^M \hat{\bf F}_\mu\delta(\hat{\bf R}_\mu-{\bf r})  
&=\sum_{\mu \nu}
\delta(\hat{\bf R}_\mu-{\bf r}) \hat{\bf G}_{\mu\nu} 
\nonumber\\
&-\nabla\lambda({\bf r})\left[
\hat{u}^{1}_{\bf r}-\hat{u}^{0}_{\bf r} +\calf'(\lambda({\bf r}))\hat{n}_{\bf r}\right]
\label{fr2}
\end{align}
Note that the last term may be written as the divergence of a tensor, because
\begin{align}
  \nonumber\\
\sum_{\mu \nu}
\delta(\hat{\bf R}_\mu-{\bf r}) \hat{\bf G}_{\mu\nu} 
&=\sum_{\mu\nu}  \hat{\bf G}_{\mu\nu} 
\frac{1}{2}\left[\delta(\hat{\bf R}_\mu-{\bf r})-\delta(\hat{\bf R}_\nu-{\bf r})\right]
\nonumber\\
&=-\nabla \hat{\boldsymbol{\Pi}}_{\bf r}
\label{gmunu}
\end{align}
where we have used the usual trick \cite{Grabert1982}
\begin{align}
\delta(\hat{\bf R}_\mu-{\bf r})-\delta(\hat{\bf R}_\nu-{\bf r})&=\int_0^1d\epsilon
\frac{d}{d\epsilon}\delta(\hat{\bf R}_\nu+\epsilon\hat{\bf R}_{\mu\nu}-{\bf r})
\nonumber\\
&=-\nabla{\hat\bf R}_{\mu\nu}\int_0^1d\epsilon
\delta(\hat{\bf R}_\nu+\epsilon\hat{\bf R}_{\mu\nu}-{\bf r})
\end{align}
where we have defined $\hat{\bf R}_{\mu\nu}=\hat{\bf R}_{\mu}-\hat{\bf R}_{\nu}$ and 
$R_{\mu\nu}=|{\bf R}_{\mu\nu}|$
and introduced {\em the virial part of the  stress tensor}
\begin{align}
  \hat{\boldsymbol{\Pi}}_{\bf r}&\equiv\frac{1}{2}\sum_{\mu\nu}  \hat{\bf G}_{\mu\nu} \hat{\bf R}_{\mu\nu} 
\int_0^1d\epsilon
\delta(\hat{\bf R}_\nu+\epsilon\hat{\bf R}_{\mu\nu}-{\bf r})
\label{Pivir}
\end{align}
In summary, we may write the force density as
\begin{align}
\hat{\bf f}_{\bf r}
&=-\nabla \hat{\boldsymbol{\Pi}}_{\bf r}
-\nabla\lambda({\bf r})\frac{\delta H^{[\lambda]}}{\delta \lambda({\bf r})} 
\label{frfin}
\end{align}
where we have used (\ref{dhdl}). As a consequence, the momentum equation (\ref{lrlg}) takes the form
\begin{align}
   i L\hat{\bf g}_{\bf r}&=-\nabla\hat{\boldsymbol{\Sigma}}_{\bf r}
-\nabla\lambda({\bf r})
\frac{\delta H^{[\lambda]}}{\delta \lambda({\bf r})} 
\label{lrlg2}
\end{align}
where the  full stress tensor  $\hat{\boldsymbol{\Sigma}}_{\bf r}=\hat{\bf
  K}_{\bf   r}+\hat{\boldsymbol{\Pi}}_{\bf  r}$   is   given  by   the
Irwing-Kirkwood (IK) form, generalized for \hadr,
\begin{align}
\hat{\boldsymbol{\Sigma}}_{\bf r}
&=\sum_\mu^M \hat{\bf P}_\mu\hat{\bf V}_\mu\delta(\hat{\bf R}_\mu-{\bf r})
\nonumber\\
&+\frac{1}{2}\sum_{\mu\nu}  \hat{\bf G}_{\mu\nu} {\bf R}_{\mu\nu} 
\int_0^1d\epsilon
\delta({\bf R}_\nu+\epsilon{\bf R}_{\mu\nu}-{\bf r})
\label{Sigma0}
\end{align}

\subsection{The temperature}
It  is worth  considering the  equilibrium average  computed  with the
canonical  ensemble  of the  kinetic  part  of  the stress  tensor  in
Eq.  (\ref{kinst}).   It  is  computed  easily   because  momentum  is
distributed according  to the Gaussian Maxwell  distribution, with the
result
\begin{align}
  \langle  \hat{\bf K}_{\bf r}\rangle^{[\lambda]}&=
k_BT\langle n_{\bf r}\rangle^{[\lambda]}{\bf  1}
\label{eqkin}
\end{align}
Closely  related to  the  kinetic part  of  the stress  tensor is  the
kinetic energy density  field of the centers of  mass which is defined
as
\begin{align}
  k_{\bf r}&\equiv\sum_\mu^M\frac{m_\mu}{2}{\bf V}_\mu^2\delta({\bf r}-{\bf R}_\mu)
\end{align}
and whose average is
\begin{align}
\langle  k_{\bf r}\rangle^{[\lambda]}
&= \frac{3k_BT}{2}\langle n_{\bf r}\rangle^{[\lambda]}
\label{prediction}\end{align}
We may introduce a CoM temperature field as the kinetic energy density
divided by the number density,  providing an idea of the local kinetic
energy of the system, through the following definition
\begin{align}
k_BT({\bf r}) &\equiv\frac{2}{3}
\frac{\langle  k_{\bf r}\rangle^{[\lambda]}}{\langle  n_{\bf r}\rangle^{[\lambda]}} =k_BT
\label{tcte}
\end{align}
where the  last identity is  just Eq. (\ref{prediction}).  This result
states  that  in  all   space  including  the  transition  region  the
temperature field is constant, $T({\bf r})=T$.

\subsection{The stress and the pressure}
The equilibrium  average of  the time rate  of change of  the momentum
density field is zero at  equilibrium, this is $\langle iL{\bf g}_{\bf
  r}\rangle^{[\lambda]}=0$ (as  can be  shown by integrating  by parts
the Liouville  operator and use of $LH^{[\lambda]}=0$).  By taking the
equilibrium average of Eq. (\ref{lrlg}) we obtain then
\begin{align}
  0&=-\nabla  \langle  \hat{\bf K}_{\bf r}\rangle^{[\lambda]}+  \langle  {\bf f}_{\bf r}\rangle^{[\lambda]}
\end{align}
which, on account of Eq. (\ref{eqkin}) gives an explicit form for the force density
field
\begin{align}
\label{fandn}
 \langle  {\bf f}_{\bf r}\rangle^{[\lambda]}&=k_BT\nabla\langle n_{\bf r}\rangle^{[\lambda]}
\end{align}
In passing, we note that Eq. (\ref{fandn}) is valid for {\em any} Hamiltonian system:
notably, this intimate relation between the force density field and the density gradients is {\em independent}
on the fluid compressibility. It explains the essence of many algorithms \cite{FritschPRL108,Issa2014,kotsalis07}
designed to impose a flat density profile by adding an external force ``correction'' to the system (which,
according to Eq. (\ref{fandn}) has to ensure vanishing total force density field ${\bf f}_{\bf r}=0$).
Figure \ref{fig:ener} (middle panel) offers a numerical check of the relation (\ref{fandn}) in one of our \hadr systems
(in that case with ${\bf f}_{\bf r} \ne 0$).
Now, let us consider the equilibrium average of Eq. (\ref{lrlg2}) by introducing
\begin{align}
  \boldsymbol{\Sigma}({\bf r})&\equiv\langle \hat{\boldsymbol{\Sigma}}_{\bf r}\rangle^{[\lambda]}
=k_BT n({\bf r})+  \boldsymbol{\Pi}({\bf r})
\nonumber\\
  \boldsymbol{\Pi}({\bf r})&=\langle \hat{\boldsymbol{\Pi}}_{\bf r}\rangle^{[\lambda]}
\label{Sigma}
\end{align}
Here   $\boldsymbol{\Sigma}({\bf   r})$   is   the  average   of   the
Irwing-Kirkwood (IK) stress tensor in Eq. (\ref{Sigma0}), which is decomposed into its ideal
and  interaction (or  excess over  ideal)  parts. With  the IK  stress
tensor, Eq. (\ref{lrlg2}) gives
\begin{align}
\nabla   \boldsymbol{\Sigma}({\bf r})=
k_BT\nabla n({\bf r})+\nabla \boldsymbol{\Pi}({\bf r})
&=-\frac{\delta F^{[\lambda]}}{\delta \lambda({\bf r})}\nabla\lambda({\bf r})
\label{crux}
\end{align}
Under  equilibrium conditions, Eq.   (\ref{crux}) just  represents the
hydrostatic balance \cite{LandauFL} i.e.  the response of the system's
equilibrium  stress field to  an external  force.  When  the switching
field is sufficiently smooth, we expect from symmetry reasons that the
average of the interaction part of the stress tensor is isotropic
 \begin{align}
 \boldsymbol{\Pi}({\bf r})&=p^{\rm ex}({\bf r}){\bf 1}
\label{Piiso}
 \end{align}
 where we have introduced the excess (over ideal) part of the pressure. The total pressure
 is defined as
 \begin{align}
   p({\bf r}) &\equiv p^{\rm id}({\bf r})+ p^{\rm ex}({\bf r})
 \nonumber\\
 p^{\rm id}({\bf r})&\equiv k_BTn({\bf r})
 \nonumber\\
 p^{\rm ex}({\bf r})&\equiv \frac{1}{3}{\rm Tr}\left[\boldsymbol{\Pi}({\bf r})\right]
 \end{align}
  Therefore, Eq. (\ref{crux})
 takes the form
 \begin{align}
 \boldsymbol{\nabla }p({\bf r})=k_BT  \boldsymbol{\nabla }n({\bf r})+  \boldsymbol{\nabla }p^{\rm ex}({\bf r})&=-\frac{\delta F^{[\lambda]}}{\delta \lambda({\bf r})}\nabla\lambda({\bf r})
 \label{Gradp}
 \end{align}

The  two exact results  (\ref{tcte}) and  (\ref{crux}) give  the local
thermodynamics of the system in  terms of its equations of state. They
are one of the main important results of the present work.
\section{Translation invariance}
\label{sec:trans}
\subsection{The free energy}
A  nice theorem  about the  free energy  involves its  behaviour under
translations.  Assume that there  are no external potential fields and
that  the   system  is  either  infinite  or   has  periodic  boundary
conditions.  We may perform  in the definition (\ref{Flam}) the change
of  variables ${\bf  r}_i={\bf r}_i'+{\bf  a}$ where  ${\bf a}$  is an
arbitrary  translation   vector.   Because  all   the  potentials  are
translational invariant, we arrive at the identity
\begin{align}
  F_{[\lambda]}&=  F_{[T_{\bf a}\lambda]}
\label{FFa}
\end{align}
where $T_{\bf a}$ is a translational operator that when applied to a function
gives
\begin{align}
  T_{\bf a}\lambda({\bf r}) &=\lambda({\bf r}+{\bf a})
\label{Ta}
\end{align}
% If there  is an  external potential $V^{\rm ext}$, the  free energy is  a functional
% also  of this external  potential. In  this case,  we have  instead of
% (\ref{FFa}) the following identity
% \begin{align}
%   F_{[\lambda,V^{\rm ext}]}&=  F_{[T_{\bf a}\lambda,T_{\bf a}V^{\rm ext}]}
% \end{align}
We may now  take the derivative of both sides  of Eq. (\ref{FFa}) with
respect to ${\bf a}$ and obtain
\begin{align}
  0&=\frac{\partial F_{[T_a\lambda]}}{\partial {\bf a}}=\int d{\bf r}\frac{\delta F_{[T_{\bf a}\lambda]}}{\delta \lambda({\bf r})}\frac{\partial}{\partial {\bf a}}T_{\bf a}\lambda({\bf r})
\label{dfdl=00}
\end{align}
where the chain rule has been used. By using (\ref{Ta}) and evaluating
the result at ${\bf a}=0$ we obtain
\begin{align}
  \int d{\bf r}\frac{\delta F_{[\lambda]}}{\delta \lambda({\bf r})}\nabla\lambda({\bf r})=0
\label{dfdl=0}
\end{align}
One  consequence  of the  translation  invariance  of  the free  energy
(\ref{FFa}) is that the average total force on the system is zero. The
average total force is
\begin{align}
  \langle {\bf F}\rangle^{[\lambda]}&=\frac{1}{Z[\lambda]}\int \frac{d^{3N}r}{\Lambda^{3N}}
\exp\{-\beta H_{[\lambda]}\}\sum_i^N\left(-\frac{\partial H_{[\lambda]}}{\partial {\bf r}_i}\right)
\nonumber\\
&=k_BT \frac{1}{Z}\int \frac{d^{3N}r}{\Lambda^{3N}}
\sum_i^N\frac{\partial }{\partial {\bf r}_i}\exp\{-\beta H_{[\lambda]}\}
\end{align}
We may again perform a translation of the origin of coordinates and produce the change
of variables ${\bf r}_i={\bf r}_i'+{\bf a}$ that becomes
\begin{align}
  \langle {\bf F}\rangle^{[\lambda]}&=
k_BT \frac{1}{Z[\lambda]}\int \frac{d^{3N}r'}{\Lambda^{3N}}
\sum_i^N\frac{\partial }{\partial {\bf r}'_i}\exp\{-\beta H_{[T_{\bf a}\lambda]}\}
\nonumber\\
&=
k_BT \frac{1}{Z[\lambda]}\frac{\partial}{\partial {\bf a}}\int \frac{d^{3N}r'}{\Lambda^{3N}}
\exp\{-\beta H_{[T_{\bf a}\lambda]}\}
\nonumber\\
&=
k_BT \frac{1}{Z[T_{\bf a}\lambda]}\frac{\partial}{\partial {\bf a}}\int \frac{d^{3N}r'}{\Lambda^{3N}}
\exp\{-\beta H_{[T_{\bf a}\lambda]}\}
\nonumber\\
&=- 
 \frac{\partial}{\partial {\bf a}}F_{[T_{\bf a}\lambda]} =0
\end{align}
where   the   last   identity   follows  from   Eq.   (\ref{dfdl=00}).
More  generally, we have derived  an important relation
  between the derivative  of the free energy functional  and the total
  force on the system,
\begin{equation}
    \langle {\bf F}\rangle^{[\lambda]} = - \int d{\bf r}\frac{\delta F_{[\lambda]}}{\delta \lambda({\bf r})} \nabla\lambda({\bf r})
\label{netforce}
\end{equation}
which indicates that $-\nabla\lambda({\bf r}) \delta F_{[\lambda]}/\delta \lambda({\bf r})$ 
is the force density field induced by the jump in potential energy densities (``the drift force'' in Ref. \cite{Potestio:PRL:110})
and the free energy correction (see Eq. \ref{dfdl=01}).
However to reach a well-defined equilibrium state, 
any prescription for computing  the free
energy compensating term entering the free energy $F_{[\lambda]}$ has
to comply with Eq. (\ref{dfdl=0}). Otherwise, a  net force (\ref{netforce})
will appear  in  the  system. 
In this sense,
the requirement (\ref{dfdl=0}) provides {\em global thermodynamic consistency}.
By integrating Eq. (\ref{crux}) over the system volume and 
using Gauss theorem, leads to
\begin{equation}
  \oint  {\boldsymbol{\Sigma}}_{\bf r} \cdot {\bf n} d r^2 = - \int  \nabla\lambda({\bf r})\frac{\delta F^{[\lambda]}}{\delta \lambda({\bf r})} d {\bf r}.
\label{gauss}
\end{equation}
Therefore   in  periodic   systems  (where   by   construction  $\oint
\boldsymbol{\Sigma}_{\bf   r}  \cdot   {\bf   n}  =0$)   translational
invariance  (\ref{dfdl=0}) and  global  thermodynamic consistency  (in
particular, mechanical  equilibrium) are trivially  satisfied for {\em
  any choice} of the free energy correction.  

\subsection{Averages of local functions}
Consider  a {\em local} function based on the CoM of the form
\begin{align}
  A_{{\bf r}}(r,p)&=\sum_\mu^M A_\mu(r,p)\delta(\hat{\bf R}_\mu-{\bf r})
\end{align}
where  $A_\mu(r,p)$  is traslationally  invariant,  so  the effect  of
changing ${\bf r}_i$ with ${\bf r}_i+{\bf a}$ for any vector ${\bf a}$
leaves $A_\mu$ invariant. 
Examples of local functions are those defined in Eqs. (\ref{defu1u0}). In this case, we have the following identity
\begin{align}
  \left\langle A_{\bf r}\right\rangle^{[T_{\bf a}\lambda]}&=  \left \langle A_{{\bf r}+{\bf a}}\right\rangle^{[\lambda]}\label{atrans}
\end{align}
\begin{widetext}
as we can check explicitly
\begin{align}
  \left\langle A_{\bf r}\right\rangle^{[T_{\bf a}\lambda]}&=  
\frac{1}{Z[T_{\bf a}\lambda]}\int d^{3N}rd^{3N}pA_{\bf r}(r,p)
\nonumber\\
&\times\exp\left\{
-\beta\left[K+V^{\rm intra}+\sum_\mu^M\lambda(\hat{\bf R}_\mu+{\bf a})V^{1}_\mu
+\sum_\mu^M(1-\lambda(\hat{\bf R}_\mu+{\bf a}))V^0_\mu+\calf(\lambda(\hat{\bf R}_\mu+{\bf a}))\right]
\right\}
\nonumber\\
&=\left\langle A_{{\bf r}+{\bf a}}\right\rangle^{[\lambda]}
\label{check}
\end{align}
where we have performed a change of variables ${\bf r}_i\to{\bf r}_i-{\bf a}$ in the last identity. 
% Note that an identical equation is obtained if the local function is based on atoms rather than
% on CoM, this is, 
% \begin{align}
%   A_{\bf r}&=\sum_i^NA_i(r,p)\delta({\bf r}_i-{\bf r})
% \end{align}
By taking  the derivative of Eq. (\ref{atrans}) with respect  to ${\bf  a}$ and
setting afterwards ${\bf a}=0$ we have
\begin{align}
\nabla \left\langle A_{{\bf r}}\right\rangle^{[\lambda]}
&=  \int d{\bf r}'\nabla'\lambda({\bf r}')\frac{\delta }{{\delta \lambda({\bf
r}')}}\left\langle A_{{\bf r}}\right\rangle^{[\lambda]}
\label{na}
\end{align}
The functional derivative of the average is given by
\begin{align}
\frac{\delta }{{\delta \lambda({\bf
r}')}}\left\langle A_{{\bf r}}\right\rangle^{[\lambda]}
% &=\frac{\delta }{{\delta \lambda({\bf
% r}')}}
% \frac{1}{Z[\lambda]}\int d^{3N}rd^{3N}pA_{\bf r}(r,p)\exp\{-\beta H_{[\lambda]}\}
% \nonumber\\
&=\beta\langle A_{\bf r}\rangle^{[\lambda]}
\left\langle\frac{\delta H_{[\lambda]}}{{\delta \lambda({\bf
r}')}}
\right\rangle^{[\lambda]}
-\beta
\left\langle A_{\bf r}\frac{\delta H_{[\lambda]}}{{\delta \lambda({\bf
r}')}}
\right\rangle^{[\lambda]}
=-\beta
\left\langle \delta A_{\bf r}\frac{\delta H_{[\lambda]}}{{\delta \lambda({\bf
r}')}}
\right\rangle^{[\lambda]}
\end{align}
where $\delta A_{\bf r}=A_{\bf r}-\langle A_{\bf r}\rangle^{[\lambda]}$.
By using (\ref{dhdl}) we obtain the exact result for local functions
\begin{align}
  \nabla  \left\langle  A_{{\bf r}}\right\rangle^{[\lambda]}  &=-\beta
  \int  d{\bf r}'\nabla'\lambda({\bf  r}') 
\left\langle  \delta A_{\bf
      r}(u^1_{{\bf    r}'}-u^0_{{\bf    r}'}+{\cal    F}'(\lambda({\bf
      r}'))n_{{\bf   r}'})  \right\rangle^{[\lambda]}
\label{nablaAr}
\end{align}  
This expression  clearly shows that  the inhomogeneities of  any local
function along space  will show up basically in  the transition region
$0<\lambda<1$ for which $\nabla  \lambda\neq0$ and are exclusively due
to  the  correlations  of  this  local function  with  the  functional
derivative of the  Hamiltonian.  For example, take the  center of mass
density field $n_{\bf r}$ as the local function $A_{\bf r}$. The above
expression gives
\begin{align}
  \nabla  \left\langle  n_{{\bf r}}\right\rangle^{[\lambda]}  &=-\beta
  \int  d{\bf r}'\nabla'\lambda({\bf  r}') 
\left[\left\langle  \delta n_{\bf
      r}(u^1_{{\bf    r}'}-u^0_{{\bf    r}'})\right\rangle^{[\lambda]}
+
{\cal    F}'(\lambda({\bf
      r}'))\left\langle  \delta n_{\bf
      r}n_{{\bf    r}'}  \right\rangle^{[\lambda]}
\right]
\label{nablan}
\end{align} 
This expression connects (linearly) the gradients of the density field
with the  gradients of the  switching function. It explains  why there
should be  molecular density variations  in the  region where  the switching
function changes its value.
\end{widetext}

\section{The free energy compensation term
  $\calf(\lambda)$ through an iterative route}
\label{sec:itera}
Up to now we have presented  a number of exact results in Eqs. (\ref{tcte}),
(\ref{Gradp}),  and  (\ref{nablaAr}),  that  are  valid  for  a  general
Hamiltonian of the form (\ref{HA}).  The particular functional form of
the free  energy compensation  term $\calf(\lambda)$  has not
yet been  specified. We will now  use these exact results  in order to
fix the functional form of the free energy compensating term.

\subsection{Constant stress field}
The basic requirement that the  free energy in the AA region coincides
with the  free energy of  the CG region, $F_{[1]}=F_{[0]}$  (i.e. that
the free energy does not depend  on the actual value of $\lambda$) can
be  generalized to  the case  that  the parameter  $\lambda$ is  space
dependent. We require that {\em  the actual free energy is independent
  of  the switching  field $\lambda({\bf  r})$}.  This  requirement is
mathematically expressed as the vanishing of the functional derivative 
\begin{align}
\frac{  \delta F_{[\lambda]}}{\delta \lambda({\bf r})}=0
\label{df0}
\end{align}
The  condition  (\ref{df0})  will   be  referred to as  the  {\em  local
  thermodynamic  consistency  requirement  of}  \hadr. Note  that  the
requirement   (\ref{df0})   ensures   automatically   the   translational
invariance of  the system expressed  in Eq. (\ref{dfdl=00}).   It also
ensures,  through  Eq.  (\ref{Gradp}),  that  the stress field and, therefore the pressure, is
constant through  space.  In general, however, the  density field will
not be constant and the  system may experience differences between the
value of  the density in the AA  region and the GG  region. Of course,
the variations of  the density are compensated with  the variations of
the excess pressure $p^{\rm ex}({\bf  r})$ in order to have a constant
pressure field.

By using Eq. (\ref{dfdl=01}), Eq. (\ref{df0}) becomes
\begin{align}
0=\left\langle {u}^1_{\bf r}-{u}^0_{\bf r}\right\rangle^{[\lambda]}+
  \calf'(\lambda({\bf r}))  \left\langle {n}_{\bf r}\right\rangle^{[\lambda]}
\label{dfdl=01b}
\end{align}
This equation can be understood as a non-linear functional equation to
be   solved   for   $\calf(\lambda)$   (where   $\calf(\lambda)$  appears  explicitly  as  well  as  implicitly  in  the
definition of  the averages $\langle\cdots\rangle^{[\lambda]}$). An iterative
method to solve Eq. (\ref{dfdl=01b}) is given in Sec. \ref{sec:simulations}.

\subsection{Constant density field }
\label{sec:nconts}
The  Hamiltonian (\ref{HA}),  with $\calf(\lambda)$ obtained
from the condition  that its free energy does not  depend on the field
$\lambda({\bf r})$ (i.e.  conditions (\ref{df0}) and (\ref{dfdl=01b})),
ensures  that the pressure  field is  constant through  the simulation
box. However,  it does not ensure  that the molecular  mass density or
the molecular energy density are the same in the AA and CG regions. We
expect that, to the  extend that the CG model is a  good model in that
it  reproduces correctly the  molecular radial  distribution function,
the density mismatch between AA and CG regions cannot be very large.

However, the CG potential is approximate and there  may be  situations  in  which  keeping the  molecular
density field  through the system  may be more important  than keeping
the  pressure field  constant.   In these  situations, an  alternative
definition   of  the   term  $\calf(\lambda({\bf   R}_\mu))$   in  the
Hamiltonian (\ref{HA}) is required.  Eq. (\ref{crux}) suggests a route
to  an  alternative  definition  of $\calf(\lambda)$  that  ensures  a
constant  density  field.   By  setting $\nabla  \langle  \hat{n}_{\bf
  r}\rangle^{[\lambda]}=0$ in Eq. (\ref{crux}) we obtain
\begin{align}
\nabla\lambda({\bf r})\frac{\delta F^{[\lambda]}}{\delta \lambda({\bf r})}
+\nabla \left\langle\hat{\boldsymbol{\Pi}}_{\bf r}\right\rangle^{[\lambda]}&=0
\label{crux0}
\end{align}
this is
\begin{align}
\nabla\lambda({\bf r})\left[
\langle\hat{u}^1_{\bf r}\rangle^{[\lambda]}-\langle\hat{u}^0_{\bf r}\rangle^{[\lambda]}+\calf'(\lambda({\bf r}))\langle\hat{n}_{\bf r}\rangle^{[\lambda]}\right]
+\nabla\langle\hat{\boldsymbol{\Pi}}_{\bf r}\rangle^{[\lambda]}&=0
\label{gibbs}\end{align}
This    equation    is   a    non-linear    implicit   equation    for
$\calf'(\lambda({\bf  r}))$  that may  be  computed  iteratively in  a
simulation   because  all  terms,   except  $\calf'$   are  explicitly
computable.  This $\calf(\lambda)$  will, by construction, ensure that
$\nabla  \langle \hat{n}_{\bf  r}\rangle^{[\lambda]}=0$, but  will not
satisfy,   in   general,   the  thermodynamic   consistency   property
(\ref{dfdl=01b}).   The pressure field  $\boldsymbol{\Sigma}({\bf r})$
will not be constant across the  system and its gradient will be given
by
\begin{align}
 \nabla \boldsymbol{\Sigma}({\bf r}) &=
-\nabla\lambda({\bf r})\frac{\delta F^{[\lambda]}}{\delta \lambda({\bf r})}
\label{nsn0}
\end{align}
where  we have  used (\ref{Sigma})  and (\ref{crux0}).   Note  that in
general,  (\ref{crux0})   does  not   comply  with  the   {\em  global
  thermodynamic consistency} requirement  (\ref{dfdl=0}) that the free
energy (\ref{Flam}) is  translationally invariant.  However, as stated
[see Eq.  (\ref{gauss})],  such requirement is automatically fulfilled
in  periodic systems,  where global  mechanical equilibrium  is always
guaranteed.

\section{The free energy compensating term $\calf(\lambda)$ through local equilibrium}
\label{sec:lea}
In this  section, we  explore the simplifications  that result  in the
calculation of  the free energy  compensation term when  the switching
field  is sufficiently  smooth in  the length  scale of  the molecular
correlations.   As formally  justified in  Appendix  \ref{localeq}, in
this case,  we may resort  to a {\em local  equilibrium approximation}
(LEA).   The  LEA essentially  consists  on  assuming  that the  
average   of   any   local microscopic   quantity   $\langle   \hat{A}_{\bf
  r}\rangle^{[\lambda]}$   obtained   from   the   \hadr   Hamiltonian
$H_{[\lambda]}$, is close to the  average where the field $\lambda$ is
constant    $\overline{\lambda}$    and    equal    to    the    value
$\overline{\lambda}=\lambda({\bf  r})$ at the  space point  ${\bf r}$.
Each value of this function determines a {\em hybrid molecular model}.
The   (canonical)   average    of   such   ``hybrid''   fluid   (using
$H_{\overline{\lambda}}$)    is    denoted    as   $\langle    \hat{A}
\rangle^{\overline{\lambda},n,T}$,  where the prescribed values of $n$
and $T$  are indicated.  If  $\lambda({\bf r})$ is smooth  enough, the
\hadr local  average at ${\bf r}$  is close to  the standard canonical
average  of a fluid  model with a constant $\overline{\lambda}=\lambda({\bf r})$
(see Eq.  Appendix \ref{localeq}),
\begin{align}
\langle \hat{A}_{\bf r}\rangle^{[\lambda]}   \approx    
\langle \hat{A} \rangle^{\overline{\lambda}=\lambda({\bf r}),
\langle n_{\bf r} \rangle^{[\lambda]}, \langle T_{\bf r} \rangle^{[\lambda]}} \equiv 
\langle \hat{A}_{\bf r}\rangle^{\lambda_{\bf r}}
\label{AAaprox}
\end{align}
where the last definition is introduced to alleviate the fully explicit
heavy notation of the local average. 
%%%

In what follows  we use the LEA expressed  in Eq. (\ref{AAaprox}) with
two purposes.  First, we derive a non-iterative route to find the free
energy  correction ${\cal  F}(\lambda)$. This  non-iterative procedure
connects  the  \hadr  formalism   to  the   process  used  in
thermodynamic integration \cite{Kirkwood1935,Frenkel.book}, from which
the \hadr idea actually stems.  Second,  we use the LEA to explore the
relations  between the  thermodynamic variables  along  the transition
region for the different forms of the free energy corrections proposed
hereby           and            in           previous           papers
\cite{Potestio:PRL:110,Potestio:PRL:111}.

\subsection{Kirkwood route to constant stress field}
\label{kirk_pres}
When  the  switching  field  varies  very smoothly,  we  may  use  the
approximation  (\ref{AAaprox})  in  Eq.  (\ref{dfdl=01}) in  order  to
obtain a method that does not require an iterative procedure. Indeed, to first order
in gradients of $\lambda({\bf r})$ we have
\begin{align}
0=\left\langle {u}^1_{\bf r}-{u}^0_{\bf r}\right\rangle^{\lambda}+
  \calf'(\lambda)  \left\langle {n}_{\bf r}\right\rangle^{\lambda}
\label{dfdl=02a}
\end{align}
where the  actual value of $\lambda$ is  $\lambda({\bf r})$. According
to the LEA, this identity can  be also understood in terms of averages
of hybrid  fluids with constant $\lambda$.  By  integrating over space
and using the definitions (\ref{defu1u0}) we obtain
\begin{align}
0&=
  \left\langle {U}^1-{U}^0\right\rangle^{\lambda}+ \calf'(\lambda)M
\label{dfdl=02}
\end{align}
where  we  have  defined   the  inter-blob  potential  energy  of  the
microscopic and CG systems as
\begin{align}
  {U}^1&=\sum_\mu^M V^1_\mu
\nonumber\\
  {U}^0&=\sum_\mu^M V^{0}_\mu
\end{align}
By integrating with respect to $\lambda$, we may write Eq. (\ref{dfdl=02}) as
\begin{align}
\calf^K(\lambda) = - \frac{1}{M} \int_0^\lambda d\lambda'\left\langle\frac{\partial U}{\partial \lambda'}\right\rangle^{\lambda'} +  {\rm C}
\label{K}
\end{align}
where  we have  defined  the potential  energy  $U\equiv\lambda U^1  +
(1-\lambda) U^0$. For consistency  with Eq. (\ref{Fm4}), the arbitrary
constant C should be set to  fix $\calf^K(1) =0$ (i.e. the free energy
correction is zero in the atomistic domain).  On  the right hand side
of  Eq.  (\ref{K}) one  recognizes the  Kirkwood formula  for standard
thermodynamic  integration  \cite{Kirkwood1935}  which indicates  that
$\calf^K(0)$ is the change  in free energy over an alchemic
transformation of the interblob interaction from $U^1$ to $U^0$.  This
is consistent  with the  interpretation given after  Eq. (\ref{F0F1}).
Evaluation of the RHS of Eq. (\ref{K}) from a series of simulations at
fixed  $\lambda$ offers a  non-iterative protocol  to the  free energy
correction $\calf$. 
Kirkwood  calibration   of  $\calf$   relies  however  on   the  local
thermodynamic equilibrium [see (\ref{AAaprox})] as Eq.  (\ref{K}) does
not ensure  the thermodynamic  consistency (\ref{df0}), except  if the
switching function  is smooth  enough.  Simulations presented  in Sec.
\ref{sec:simulations}  show that  in  practice Kirkwood  non-iterative
approximation works quite well, at least for the test cases considered
here. This  was also observed  in previous works with  different fluid
models \cite{Potestio:PRL:110,Potestio:PRL:111},  although a study of
the  validity of Kirkwood  TI as  a function  of the  transition layer
length  and the  coupled fluid  models  was not  considered.  We  will
perform such study in Sec. \ref{sec:simulations}.

\subsection{Kirkwood route to constant density field.}
\label{kirk_dens}
We now consider the local equilibrium approximation (LEA) 
to find a non-iterative way to compute the free energy compensation term when
the target is to keep the density field constant across the simulation
box. The exact result in Eq. (\ref{crux}) can be written as
\begin{eqnarray}
  k_BT \boldsymbol{\nabla }\langle \hat{n}_{\bf  r}\rangle^{[\lambda]}+  \boldsymbol{\nabla }
  \left\langle\hat{p}^{\rm ex}_{\bf r}\right\rangle^{[\lambda]}
  +\frac{\delta F^{[\lambda]}}{\delta \lambda({\bf r})}\nabla\lambda({\bf r})&=&0 
  \nonumber\\
  k_BT \boldsymbol{\nabla }\langle  \hat{n}_{\bf  r}\rangle^{[\lambda]}+  \boldsymbol{\nabla }
  \left\langle\hat{p}^{\rm ex}_{\bf r}\right\rangle^{[\lambda]} +\left[\langle\hat{u}^1_{\bf r}-\hat{u}^0_{\bf
      r}\rangle^{[\lambda]}+\calf'(\lambda)\langle\hat{n}_{\bf
      r}\rangle^{[\lambda]}  \right]\nabla\lambda({\bf r})&=&0
  \label{Gradp2}
\end{eqnarray}
where the microscopic excess pressure is defined by
\begin{align}
\hat{p}^{\rm ex}_{\bf r} = \frac{1}{3}{\rm Tr}\left[\boldsymbol{\Pi}_{\bf r}  \right]
\end{align}
We  assume  that  $\langle  \hat{n}_{\bf  r}\rangle^{[\lambda]}=n$  is
constant  and,  therefore,  the   first  term  in  Eq.  (\ref{Gradp2})
vanishes.  The second  term,  with the  local equilibrium  approximation
(\ref{AAaprox}), becomes
\begin{align}
  \boldsymbol{\nabla }
\left\langle\hat{p}^{\rm ex}_{\bf r}\right\rangle^{[\lambda]} &\approx
  \frac{d}{d\lambda}
\left.   \langle       \hat{p}^{\rm ex}_{\bf r}\rangle^{\lambda}\right|_{\lambda=\lambda({\bf   r})}
\nabla \lambda({\bf r})
\end{align}
The term involving the difference between potential energy densities is,
under the local equilibrium approximation (\ref{AAaprox})
\begin{align}
  \left \langle\hat{u}^1_{\bf r}-\hat{u}^0_{\bf r}\right\rangle^{[\lambda]}\approx
\left.  \left \langle\hat{u}^1_{\bf r}-\hat{u}^0_{\bf r}\right\rangle^{\lambda}\right|_{\lambda=\lambda({\bf r})}
\end{align}
This
 may be written as 
a total derivative with respect to $\lambda$ as
\begin{align}
  \left \langle\hat{u}^1_{\bf r}-\hat{u}^0_{\bf r}\right\rangle^{\lambda}
&=\frac{d}{d\lambda}\int_0^\lambda d\lambda'
  \left \langle\hat{u}^1_{\bf r}-\hat{u}^0_{\bf r}\right\rangle^{\lambda'}
\end{align}
By collecting these last results, Eq. (\ref{Gradp2}) becomes
\begin{align}
&\left. \frac{d}{d\lambda} \left[  
 \langle       \hat{p}^{\rm ex}_{\bf r}\rangle^{\lambda}
+\int_0^\lambda d\lambda'
  \left \langle\hat{u}^1_{\bf r}-\hat{u}^0_{\bf r}\right\rangle^{\lambda'}
+\calf(\lambda) n  \right]\right|_{\lambda=\lambda({\bf   r})}
\nonumber\\
&\times\nabla\lambda({\bf r})=0
\end{align}
One way  to ensure  this identity and,  therefore, a  constant density
field through the system is by requiring
\begin{align}
   \langle       \hat{p}^{\rm ex}_{\bf r}\rangle^{\lambda}
+\int_0^\lambda d\lambda'
  \left \langle\hat{u}^1_{\bf r}-\hat{u}^0_{\bf r}\right\rangle^{\lambda'}
+\calf(\lambda) n  &= {\rm C}
\label{req1}
\end{align}
where C is a constant.
Because the averages are performed with a constant switching field, we
have  translation  invariance and  we  can  get  rid off  the  position
dependence by simply averaging (\ref{req1}) over the whole volume. This gives
\begin{align}
   \langle       \hat{P}^{\rm ex}\rangle^{\lambda}
+\frac{1}{V}\int_0^\lambda d\lambda'
  \left \langle\hat{U}^1-\hat{U}^0\right\rangle^{\lambda'}
+\calf(\lambda) n  &= {\rm C}
  \end{align}
where 
\begin{align}
  \hat{P}^{\rm ex}&\equiv\frac{1}{V}\int d{\bf r}\hat{p}_{\bf r}^{\rm ex}
=\frac{1}{V}\frac{1}{6}\sum_{\mu\nu}  \hat{\bf G}_{\mu\nu}\esc \hat{\bf R}_{\mu\nu} 
\end{align}
where  we have  used (\ref{Piiso})  and (\ref{Pivir}).  Therefore, the
non-iterative  prescription  for the  free  energy compensating  term,
valid for  smooth switching fields,  that produces a  constant density
field is
\begin{align}
\calf^K(\lambda)   &= -\frac{1}{M}\int_0^\lambda d\lambda'
  \left \langle\hat{U}^1-\hat{U}^0\right\rangle^{\lambda'}
-\frac{  \langle       \hat{P}^{\rm ex}\rangle^{\lambda}}{n} + 
{\rm C}
\label{calFcdens}
\end{align}
to be  compared with the prescription (\ref{K})  that produces a
constant pressure  field. Again, the constant  C should be  set to fix
$\calf(1)=0$.  The  non-iterative calibration of $\calf$  based on Eq.
(\ref{calFcdens})    involves    a    series    of    simulations    of
constant-$\lambda$  fluids in  the  canonical ensemble  at the  target
density $n=M/V$  and temperature  $T$.  The first  term in the  RHS of
Eq.  (\ref{calFcdens}) is  then  the  difference  in the  Helmholtz
excess free energy  (per particle) $f^{ex}(0)-f^{ex}(\lambda)$ between
the  CG  fluid  model  ($\lambda=0$)  and a  fluid  model  with  fixed
$\lambda$.  The  free energy correction $\calf$ acts  like an external
potential field  in the system  so the system's chemical  potential is
\cite{LandauSP}   $\mu    =   g(\lambda)   +    \calf(\lambda)$   where,
$g(\lambda)=f(\lambda) +  p/n$ is the Gibbs free  energy per particle,
containing ideal  and excess parts $g=g^{id}(n)  +g^{ex}$. At constant
density, the ideal part  contribution of any thermodynamic function is
constant and Eq. (\ref{calFcdens}) can be written as,
\begin{align}
g(\lambda)+\calf(\lambda) = g(1)=\mu,
\label{mu0}
\end{align}
showing that the constant density \hadr 
consistently provides a constant chemical potential $\mu$
over the system.

\section{Simulations}
\label{sec:simulations}

This   section  presents  molecular   dynamics  (MD)   simulations  to
illustrate and validate  the \hadr theoretical framework.  Simulations
of   the  microcanonical   ensemble  of   the  \hadr   Hamiltonian  in
Eq.   (\ref{H1})  were   done  in   periodic  boxes   with  dimensions
$L_x,L_y=L_z$.    We   have    used   the   tetrahedral   fluid   model
\cite{praprotnik:224106,Matej_JCP07,rdb08,Potestio:PRL:110}  which has become  one of  the benchmark
models  for Adaptive  Resolution.  Each  tetrahedral  molecule contains
four  atoms bonded  by  FENE potentials.  Non-bonded interactions  are
described  by a  purely repulsive  Lennard-Jones potential  (cutoff at
$r_{cut}=2^{1/6}\sigma$  (where $\sigma$  is the  atomic LJ-diameter).
The  coarse-grained   potential  used  for   $\lambda=0$  (CG  domain)
corresponds     to     the     Morse     potential     proposed     in
Ref. \cite{praprotnik:224106,rdb08}
\begin{equation}
  U_{cg}(r)= \gamma \left(1.0-\exp\left[-\kappa (r-r_0)\right]\right)^2
\end{equation}
The parameters $\gamma=0.105$, $\kappa=2.4$ and
$r_0=2.31$, were  originally fitted so  as to correctly  reproduce the
molecular radial distribution function of the polyatomic fluid and its
pressure. In order to study the flexibility of H-AdResS to compensate for
free energy differences between the coarse-grained and atomistic model  we 
have tweaked the CG potential to consider two cases,
\begin{itemize}
\item Fitted CG:  $\gamma=0.105$, $\kappa=2.4$ and $r_0=2.31$,
\item Non-fitted CG: $\gamma=0.305$, $\kappa=2.4$ and $r_0=2.31$
\end{itemize}
The  Inverse  Boltzmann  procedure  was  used to  set  the  fitted  CG
potential for  a molecular  density $\rho_m =0.1  \sigma^{-3}$ (atomic
density $n=4\rho_m$) and  temperature $T\simeq 1.0 \epsilon/K_B$.  The
CG potential also ensures $p^0(n,T) = p^1(n,T)$.
We consider  a simple \hadr  set-up where the switching  function only
depends on the  $x$-coordinate, $\lambda=\lambda(x)$ 
and its gradient is directed in $x$-direction, $\nabla  \lambda({\bf r})=\lambda'(x){\bf e}_x$.
%the vector Eq. (\ref{gibbs}) has only one non-vanishing component, along the  $x$ 
%\footnote{Simulations  with rigid walls and non-periodic  $\lambda(x)$ were also performed,  but not discussed  hereby.}. 
The resolution function $\lambda(x)$ is $\lambda=1$ at the AA domain and $\lambda=0$ at the CG domain
while in the transition layer it varies like,
\begin{equation}
\label{lambda}
\lambda(x)=\cos^2\left[\frac{\pi}{2}\frac{x-x_1}{\lh}\right]
\end{equation}
with $\lh=|x_1-x_0|$ the width of the transition region, where $\lambda'\ne 0$.
Here $x_1=x(\lambda=1)$ is the position of the AA-HYB border
and $x_0$ the location of the $\lambda=0$ border.

\subsection{Basic equilibrium thermodynamics of \hadr}
\begin{figure}[t]
  \includegraphics[scale=0.8]{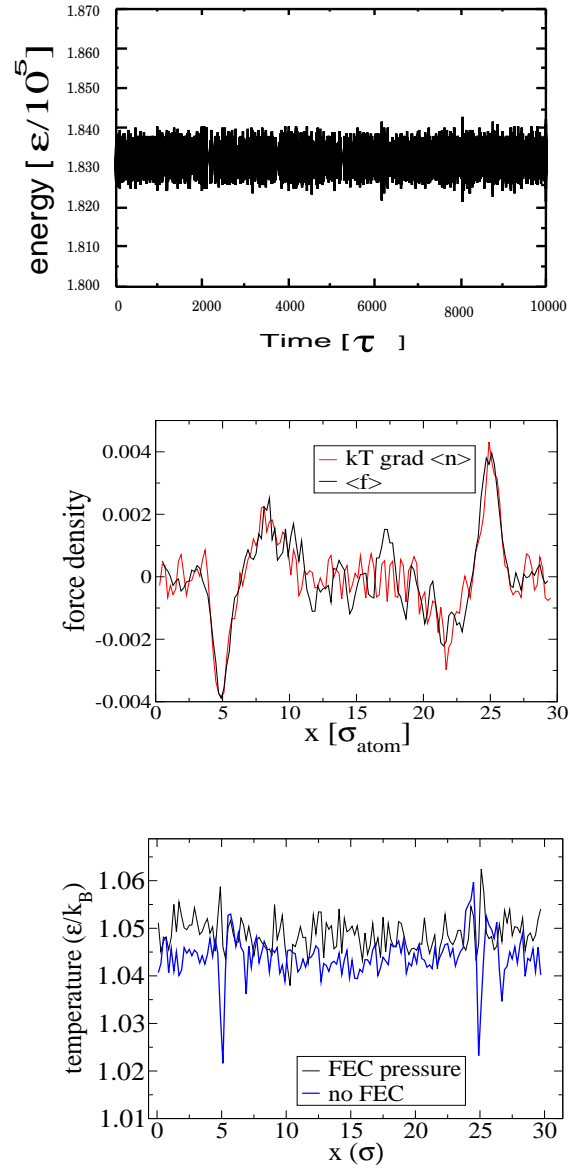}
  \caption{Top panel: Energy of  a H-Adress  simulation. There is  practically no
    drift  in total energy  over long  simulation runs  (here $5\times
    10^5  \tau$,  with  $\tau=\sigma \sqrt(m/\epsilon)$  the  standard
    Lennard-Jones  time unit  of  the atomic  potential). Middle panel: A  numerical
    cross-check of the relation (\ref{fandn}). Bottom panel: The temperature profile
    over   the    system.    Simulations   were    done   at   density
    $n=0.4\sigma^{-3}$ with fitted CG-potentials.}
\label{fig:ener}
\end{figure}

The MD algorithm was  implemented in single precision arithmetic using
a standard  second order velocity-Verlet integrator and  a Verlet list
for  neighbours search.  As  shown in  Fig. \ref{fig:ener} (top panel)  the total
energy is  conserved (up  to about $0.1\%$  deviation) and  the energy
drift  over long  runs  is practically  zero.  Figure  \ref{fig:ener} (middle panel)
illustrates  a  numerical  cross-check  of  the  interesting  relation
(\ref{fandn}),  that  relates  the  force  density  with  the  density
gradient (in the figure, for a system without free energy correction).

Also,  in Fig.  \ref{fig:ener} (bottom panel)  the temperature  profiles obtained  in
several  type of  \hadr simulations  (with or  without  correction) is
presented.  In all cases,  thermal equilibrium is attained and ensures
a  constant   temperature  profile   over  the  simulation   box.   In
microcanonical simulations the temperature  is not an input simulation
parameter so  one should expect  small variations in  temperature upon
inclusion of some form of the free energy correction (see for instance
Fig.   (\ref{fig:ener}). In  fact, a modification of the 
FEC term  changes  the overall Hamiltonian  of  the system  and  in  general  its second  derivatives
(e.g. the  heat capacity) determining  the caloric equation  of state.
For this reason, here we use a standard  (canonical)
thermostat {\em while adjusting} the free energy compensation in the 
iterative way.

\subsection{Iterative evaluation of the free energy correction}

The iterative evaluation of the  free energy correction (FEC) is based
on  the force  balance  in  Eq. (\ref{crux}),  where  the free  energy
derivative  is given  by  Eq.  (\ref{dfdl=01}).   The virial  pressure
gradient   in   Eq.     (\ref{crux})   stems   from   the   inter-blob
forces. Instead  of evaluating its  gradient, it is more  efficient to
use Eq.   (\ref{gmunu}).  We assume that the  field $\lambda({\bf r})$
changes only  along the $x$ axis,  i.e.  $\lambda({\bf r})=\lambda(x)$
and that  there is translation invariance  along the $y,z$  axis due to
the periodic boundary conditions.  This allows to average (\ref{crux})
with respect $y,z$. We introduce the following $x$ dependent fields
\begin{align}
g(x)&\equiv \langle \nabla \hat{\boldsymbol{\Pi}}_{\bf r}\cdot {\bf e}_x\rangle^{[\lambda]} = 
\left\langle\sum_{\mu \nu} \delta(\hat{X}_\mu-x) \hat{\bf G}_{\mu\nu} \cdot {\bf e}_x \right \rangle^{[\lambda]}\nonumber\\
{u}^1(x)-{u}^0(x) &\equiv\left\langle\sum^M_\mu \left(V^1_\mu -V^{0}_\mu\right) \delta(\hat{X}_\mu-x)\right\rangle^{[\lambda]}
\nonumber\\
n(x)&\equiv\left\langle\sum^M_\mu\delta(\hat{X}_\mu-x)\right\rangle^{[\lambda]}
\label{defu1u0x}
\end{align}
The  density field  $a(x)$ of  any microscopic  quantity  $A_{\mu}$ is
numerically  evaluated  by  a   binned  Dirac  delta:  $\delta_h(r)  =
\Theta_h(r)/V_{h}$  where $V_{h}$  is the  volume of  the bin  and in 1D the
characteristic  function is $\Theta_h(x)=1$  if $|x|  \le h/2$  and zero
otherwise.   As  customary  we  assume  ergodicity  and  use  temporal
averages instead of ensemble averages
\begin{equation}
  a(x) = \frac{1}{T_{\rm sample}} \int_{T_{\rm sample}} dt \sum_{\mu} A_{\mu}(t) \delta_{\Delta x} \left(x-x_{\mu}\right).
\end{equation}
The sampling time is $T_{\rm sample}$ and the volume of the bin is
$V_{\Delta x}  = \Delta  x L_y\, L_z$  with $L_{\alpha}$  the
system's size in $\alpha$ direction.

With  the  definitions  (\ref{defu1u0x}), the  mechanical  equilibrium
equation Eq. (\ref{crux}) becomes in the 1D setting
\begin{align}
\ffec(x) =& \frac{u^1(x)-u^0(x)}{n(x)} \lambda^{\prime}(x)
\nonumber\\
&+ \frac{g(x)}{n(x)} -k_BT \frac{d \ln n(x)}{dx} 
\label{gibbs2}
\end{align}
where we have introduced the ``compensation'' force
\begin{align}
\ffec(x)\equiv-  \calf'(\lambda(x))\lambda^{\prime}(x)
\end{align}
As it is clear from Eq. (\ref{eqmot1}), this is the $x$ component of the force due
to the FEC acting on the atoms of the system when they have the $x$ coordinate.

Eq. (\ref{gibbs2}) is valid  for \textit{any} form of
the FEC  $\calf(\lambda)$ as it  reflects the condition  of mechanical
equilibrium.    The prescription to have a constant pressure field
in all the system, i.e. Eq. (\ref{dfdl=01b}), becomes in the 1D setting
\begin{align}
\ffec(x)&= \frac{u^1(x)-u^0(x)}{n(x)} \lambda^{\prime}(x)
\label{1Dpconst}
\end{align}
while the condition of constant density field, Eq. (\ref{gibbs}),  becomes
\begin{align}
\ffec(x)&= \frac{u^1(x)-u^0(x)}{n(x)}  \lambda^{\prime}(x)
+ \frac{g(x)}{n(x)}
\label{1Dnconst}
\end{align}
Note  the the  fields $n(x),u^0(x),u^1(x),g(x)$  depend  implicitly on
$\calf(\lambda)$  because  they  are  given in  terms  of  equilibrium
averages computed  with a Hamiltonian  that contains $\calf(\lambda)$.
Therefore,  we  need to  solve  (\ref{1Dpconst}) and  (\ref{1Dnconst})
iteratively.  The general structure of Eqs. (\ref{1Dpconst}),(\ref{1Dnconst}) is
\begin{align}
  \ffec=\Phi(\ffec)
\end{align}
One way to solve this equation iteratively is 
\begin{align}
  F_{\mathrm{c}}^{n+1}=\Phi(F_{\mathrm{c}}^n)
\label{a1}
\end{align}
with some initial good guess  $\ffec^0$. In the present case, the Kirkwood
estimate  for $\calf(\lambda)$  is a  good  guess that  allows to  use
(\ref{a1}). If we do not have such a good initial estimate, we need to
change  the atomic  forces $\ffec(x)$  slowly, otherwise  the abrupt
change  in  the  forces  on  the  atoms may  lead  to  undesirable
perturbations such as heat production (here we use thermostats 
{\em only} during the FEC calibration), density
waves (that in a periodic system  take a long time to be adsorbed), or
even the system  explosion.  For this reason it  is better to consider
the iterative protocol
\begin{align}
\ffec^{n+1}=\ffec^n+\alpha(\Phi(\ffec^n)-\ffec^n)
\label{relax_eq}
\end{align}
where $\alpha$ is sufficiently small. When convergence is reached $\ffec^{n+1}\approx \ffec^n$ implying
$\ffec^n\approx\Phi(\ffec^n)$.
Note that
$\alpha$ can be seen as the inverse of a relaxation time (the solution
ideally  converging  exponentially  fast  to the  converged  solution,
$\ffec^{n+1}  = \ffec^n$).  We  update Eq.   (\ref{relax_eq}) each
sampling interval $T_{sample}=N_s \Delta t$ (with $N_s \sim 10^3$ time
steps)             and             in            such             case
$\alpha=\widehat{\alpha}\,\delta_{Kr}\left[\mathtt{mod}(n,N_f);0\right]$,
where $\delta_{Kr}$  is the  Kronecker delta, $n$  is the  time step,
$\mathtt{mod}(n;m)$ is the modulus function and $\hat{\alpha}<1$.

The iterative solution of the constant pressure FEC equation (\ref{1Dpconst}) becomes now
\begin{align}
F_{\mathrm{c}}^{n+1}(x)&=F_{\mathrm{c}}^{n}(x)
\nonumber\\
&+\alpha\left(\left[ \frac{u^1(x)-u^0(x)}{n(x)}\right]^n \lambda^{\prime}(x)-F_{\mathrm{c}}^n(x)\right)
\label{1DpconstIter}
\end{align}
where the notation $[\cdots]^n$ means that all averages are computed with the force $F_{\mathrm{c}}^n(x)$ known at the $n$-th iteration.

The  iterative   solution  of   the  constant  density  FEC  equation
(\ref{1Dnconst}) requires  a further  step in order  to have  a faster
convergence rate.  The idea  is to first  perform an iteration  of the
type (\ref{relax_eq})
\begin{align}
  F^*_{\mathrm{c}}&=F^{n}_{\mathrm{c}}(x)
+\alpha\left(\left[ \frac{u^1(x)-u^0(x)}{n(x)}\right]^n \lambda^{\prime}(x)\right.
\nonumber\\
&+\left.
\left[\frac{g(x)}{n(x)}\right]^n-F^n_{\mathrm{c}}(x)\right)
\label{F*}
\end{align}
Then, we iterate the equivalent condition $k_BT\nabla\ln n(x)=0$
\begin{align}
  F_{\mathrm{c}}^{n+1}(x)&=  F_{\mathrm{c}}^{n}(x)+\alpha k_BT\frac{d}{dx}\ln n(x)
\end{align}
which we further integrate over the hybrid layer to have
\begin{align}
\label{intcal}
  \calf^{n+1}(0)&=  \calf^n(0)+\alpha k_T\ln \frac{n(x_1)}{n(x_0)}
\end{align}
where we have introduced 
\begin{align}
\label{intcal2}
    \calf^n(0)&\equiv\int_{x_0}^{x_1}dxF^n_{\mathrm{c}}(x)
\end{align}
and finally correct the result (\ref{F*}) as
\begin{align}
  F_{\mathrm{c}}^{n+1}(x)&=F_{\mathrm{c}}^*(x)\frac{\calf^{n+1}(0)}{\calf^{n}(0)}
\label{Fcorrect}
\end{align}
The step in Eq. (\ref{intcal}) involves the integral (\ref{intcal2}) over the transition layer 
so it permits to substantially reduce the fluctuations of the (total) free energy jump estimation
[$\calf(0)$]. This fastens up the iterative evaluation of the compensation force $F_{\mathrm{c}}(x)$.
An analysis of the convergence rates is however left for future work.

\subsection{Fitted CG potentials}
\label{fittedcg}
\subsubsection{Kirkwood TI versus iterative evaluation of $\calf$:
the effect of hybrid layer width $\lh$.}

This section analyzes the  dependence of $\calf(\lambda)$ on the width
$\lh$  of the  transition layer.   Results will  be compared  with the
Kirkwood  thermodynamic  integration  $\calf^K(\lambda)$  whose  value
$\calf^K(0)$ at $\lambda=0$ is the free energy difference between both
fluid models  (CG and AA).  Recall that  by construction $\calf(1)=0$,
and that  for fitted CG potentials,  by definition of  fitted, we have
that $\calf(0)=0$. At some $0<\lambda(x)<1$, the agreement between the
Kirkwood free  energy $\calf^K(\lambda)$ and  the iterative evaluation
of  $\calf(\lambda)$   will  indicate   the  validity  of   the  local
equilibrium  approximation  introduced  in  Sec.   \ref{sec:lea}.  
For large enough CG and AA domains the value of $\calf(0)$
has to be independent on the width of the transition layer.

The optimal result would be $\calf(0)=\calf(0)^{K}$ for any $\lambda$,
(i.e. for any  width $\lh$). Such result would  allow the \hadr scheme
to act  as a flexible and  efficient tool for  free energy differences
evaluation.   Although  we  will  not  focus here  on  this  important
thermodynamic aspect of \hadr, we  will analyze the effect of $\lh$ on
$\calf$   by   considering   systems   with   fitted   CG   potentials
($\calf(0)=\calf(1)=0$) in constant pressure \hadr simulations.  These
issues will  be also considered later when  analyzing constant density
\hadr under non-fitted potentials, $\calf(0)\ne 0$.

The convergence of $\calf'$  is particularly fast in constant pressure
simulations because it only  involves averages of extensive quantities
(energies).  To get enough  statistics for $\calf'$ in each iteration,
$T_{\rm sample}$  can be chosen  to be few molecular  collision times.
We    usually   started   the    iterative   FEC    evaluation   using
$\calf(\lambda)=0$ as  starting seed which is certainly  a benefit, as
it avoids the pre-evaluation of  the Kirkwood free energy $\calf^K$ as
starting  point for  the  iterative route.   It  has to  be said  that
Molecular  Dynamics \hadr  only  requires the  derivative  of the  FEC
$\calf'$ for time stepping.  In this context, MD-\hadr 
\cite{Potestio:PRL:110} offers a benefit over
Monte Carlo \hadr \cite{Potestio:PRL:111}  because it permits to use a
force balance like Eq. (\ref{1Dpconst})  to iteratively evaluate/update
the FEC on-the-fly.

\begin{figure}[t]
  \includegraphics[scale=0.6]{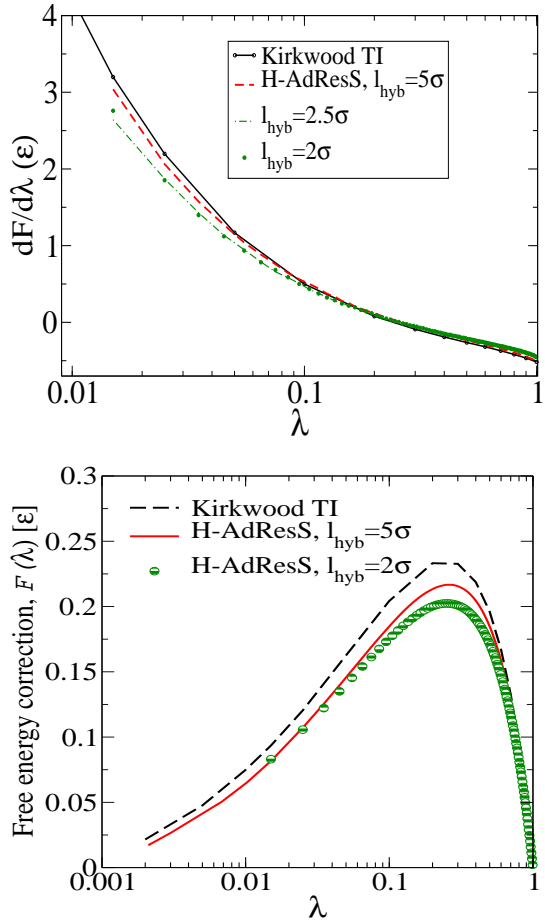}
\caption{The  derivative  $\calf'$   (top)  and  FEC  $\calf(\lambda)$
  (bottom) between  the atomistic tetrahedral  fluid and the  fitted CG
  model as  a function of $\lambda$ in  constant pressure simulations.
  Comparison  is  made  between  the  Kirkwood TI  (\ref{K})  and  the
  iterative solution of  Eq.  (\ref{gibbs2}) for several transition
  layer  widths  $\lh$.}
\label{kirk}
\end{figure}

Fig.  \ref{kirk} compares
the Kirkwood  approximation to $\calf$ with the  iterative solution of
Eq.  (\ref{gibbs2})  in a case  with $\lh=5\sigma$.  For  large enough
transition  layers,  molecular   correlations effects lessen  and  we  expect
$\calf'$ to  approach to Kirkwood's  value.  To analyze  how molecular
correlations affect $\calf'$  we have reduced the width  of the hybrid
layer  $\lh$ up  to  quite small  values.  Fig. \ref{kirk}  presents
results for  $\lh=2$, $2.5$ and $5  \sigma$, which are  similar to the
molecules'  diameters (about  $2.5\sigma$).   Remarkably, ${\cal  F'}$
becomes  quite close  the Kirkwood  free energy  as soon  as  $\lh$ is
larger  than  about twice  the  molecular  cutoff  radius.  Maybe  not
unexpectedly, deviations  between the iterative  $\calf$ and $\calf^K$
(Kirkwood)  increase  around  $\lambda=0$  and  $\lambda=1$.   Despite
differences in  $\calf(\lambda)$, it is  important to stress  that for
any choice of $\lh$ (see Fig. \ref{kirk}b) the iterative evaluation of
$\calf$ correctly predicts $\calf(0)=\calf(1)$. We shall come back to this
later in the case of non-fitted potentials.

\begin{figure}[t]
  \includegraphics[scale=0.35]{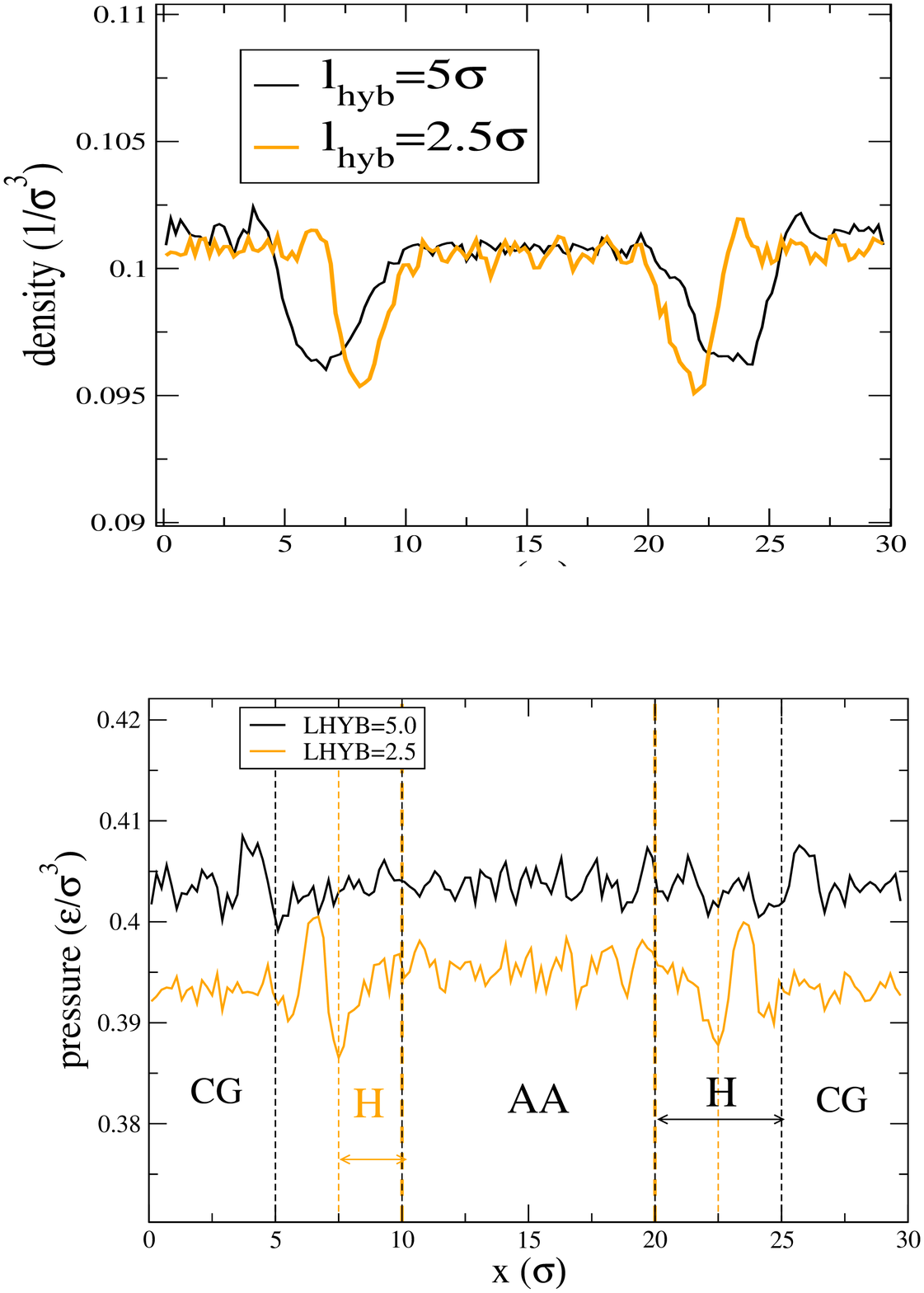}
\caption{
The effect  of reducing $\lh$ in the density and pressure profiles.
Results corresponds to fitted CG potential at constant \hadr pressure.}
\label{lhyb}
\end{figure}

Fig.   \ref{lhyb} illustrates  the  effect of  reducing  $\lh$ in  the
density  and  pressure profiles  in  \hadr  simulations with  constant
pressure.    An  interesting   observation   is  that   the  jump   of
not-compensated  quantities over the  transition layer  (here density)
does  not significantly  increase as  $\lh$  is made  shorter.  It  is
important to  notice that in a  closed system, any  mass difference in
the   transition  layer   (which  is   a  lower   density   region  in
Fig. \ref{lhyb}) induces finite  size effects.  The mass excluded from
the  transition  domain  is  transferred  to the  CG  and  AA  domains
(according to their  local chemical potential) so the  density in both
domains  will increase  over the  mean value  $\bar{n}=M/V$  (which is
indicated with  a dashed  line in Fig.   \ref{lhyb}a).  Paradoxically,
for this reason the density profile using $\lh=2.5\sigma$ is closer to
$\bar{n}$  than  the  profile  using from  $\lh=5.0\sigma$  (see  Fig.
\ref{lhyb}a).  This mismatch in the bulk densities is reflected in the
total pressure, whose (constant)  value slightly depends on $\lh$ (see
Fig. \ref{lhyb}b).

\subsubsection{Other finite box effects in closed systems}
\begin{figure}[t]
\includegraphics[scale=0.4]{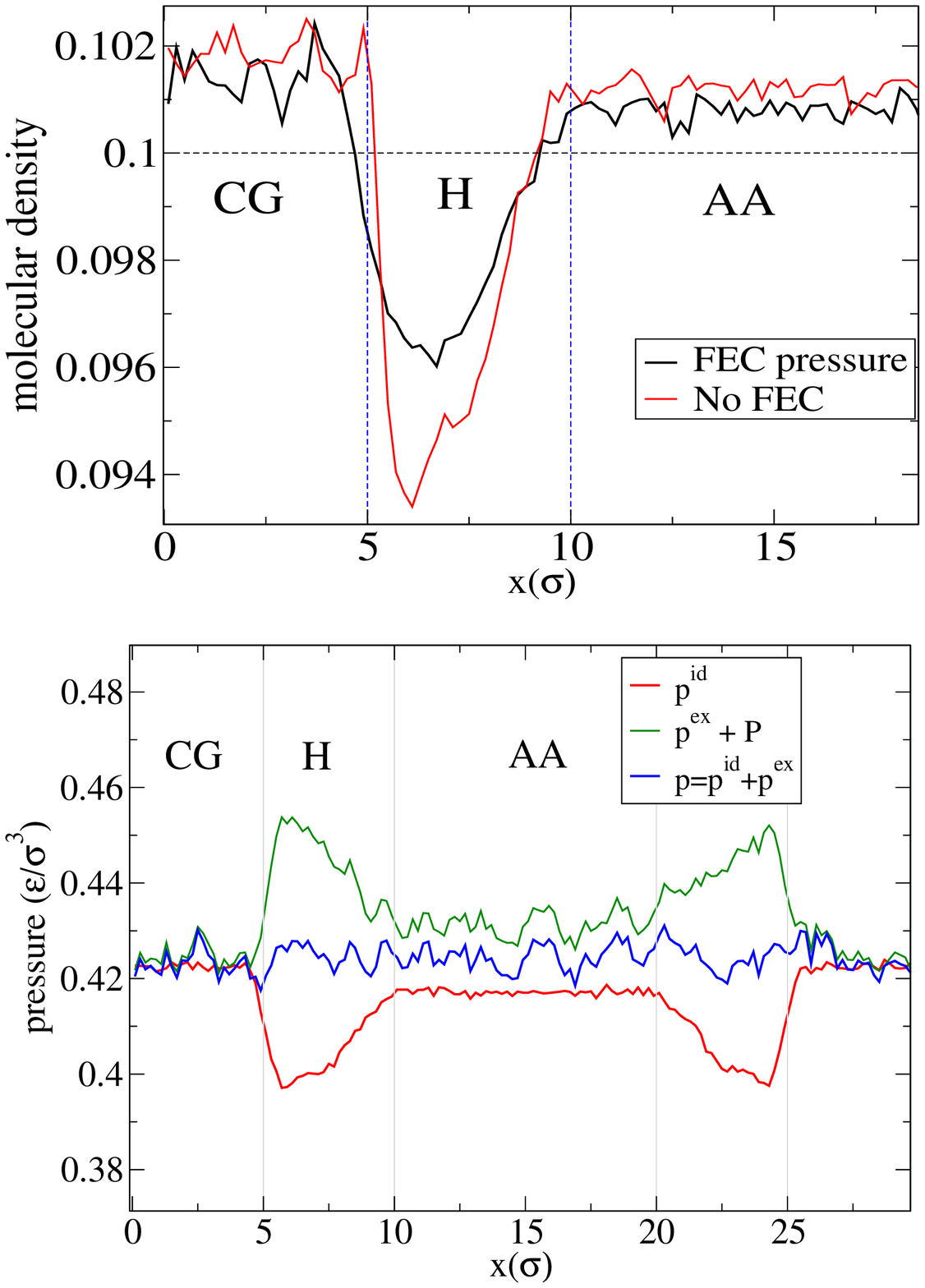}
\caption{(a) Density profile obtained from \hadr with fitted CG potentials
with constant pressure FEC and without FEC. 
(b) The kinetic $p^{id}$ and virial $p^{ex}$ contributions to the total
pressure $p=p^{id}+p^{ex}$ in the pressure compensated \hadr with fitted CG potential. The mean pressure is $P=(1/L) \int_0^L p(x) dx$. 
\label{fitcg}
}
\end{figure}

Fig.  \ref{fitcg}  shows  the  density  and  pressure  profiles  for
$\lh=5\sigma$ in the case of  fitted CG potentials. Comparison is made
between simulations with $\calf$  given by the pressure correction Eq.
(\ref{gibbs2}) and  with $\calf=0$. Some conclusions  can be extracted.
First the non-compensated version
presents  a  larger  density  jump  over the  transition  regime,  when
compared  with the  pressure compensated  \hadr.  The  overall density
mismatch  across  the transition  region  is  slightly  larger in  the
non-compensated \hadr, although it
is  not a  large difference  neither.  Second,  in closed  boxes (here
periodic) a rarefied transition  region induces finite size effects on
the bulk densities which become larger than $\bar{n}=M/V$.  The effect
is  larger for  $\calf=0$, although  this effect  is observed  in both
simulations.  This brings about consequences in the kinetic and virial
pressure profiles, shown in  Fig.  \ref{fitcg}b.  Notably, the kinetic
pressure  $p^{id}=\langle k_x\rangle$  is equal  to $k_B  T\langle n_x
\rangle$ (see Eq. \ref{tcte}) so any mismatch in density is reproduced
in $p^{id}$. The total pressure $p(x)=p^{id}(x)+p^{ex}(x)$ is robustly
fixed to a constant value  $p(x)=P$ by the FEC.  Consequently $p^{ex}$
compensates any variation in $p^{id}$ across the transition layer.

\subsection{Non-fitted CG potentials}

We  now  explore one  of  the  main benefits  of  \hadr  which is  the
possibility of working with non-fitted CG potentials.  This benefit is
not  only to  alleviate the  time consuming  and  computational effort
related to pre-evaluation  of CG potentials.  In fact,  fitting the CG
potential is a good practice as we have already seen that it minimizes
the mismatch in non-fitted thermodynamic variables. The benefits arise
from the possibility of performing simulations involving thermodynamic
{\em  processes}, which  involve changes  in the  global environmental
variables (temperature, pressure, chemical potential).  In these cases
\hadr  permits  to  work with  a  single  CG  model whose  $\calf$  is
self-adapted  over  the  whole  process  to keep  the  desired  global
constraint (pressure,  density, etc).  In  this sense \hadr  offers an
alternative to the (probably  more involved) problem of {\em potential
  transferability}. Other benefits to be considered are the evaluation
of  free  energies  differences  in  systems  involving  large  solute
molecules.  For  these applications the estimation of  the {\em total}
free  energy difference  between (CG  and  AA) models  should be  {\em
  independent on the  choice of the hybrid layer}  and should coincide
with the  Kirkwood thermodynamic value.   On the other hand  we expect
that the iterative evaluation of  $\calf'$ will reduce or suppress the
oscillations  in  the  density   (or  pressure)  profiles  around  the
transition layer.  As stated  around Eq. (\ref{nablan}), these are due
to  molecular correlations and  have been  reported in  Kirkwood based
pre-evaluated             FEC             corrections             (see
e.g. \cite{Potestio:PRL:110,Potestio:PRL:111}).

\begin{figure}[t]
  \includegraphics[scale=0.35]{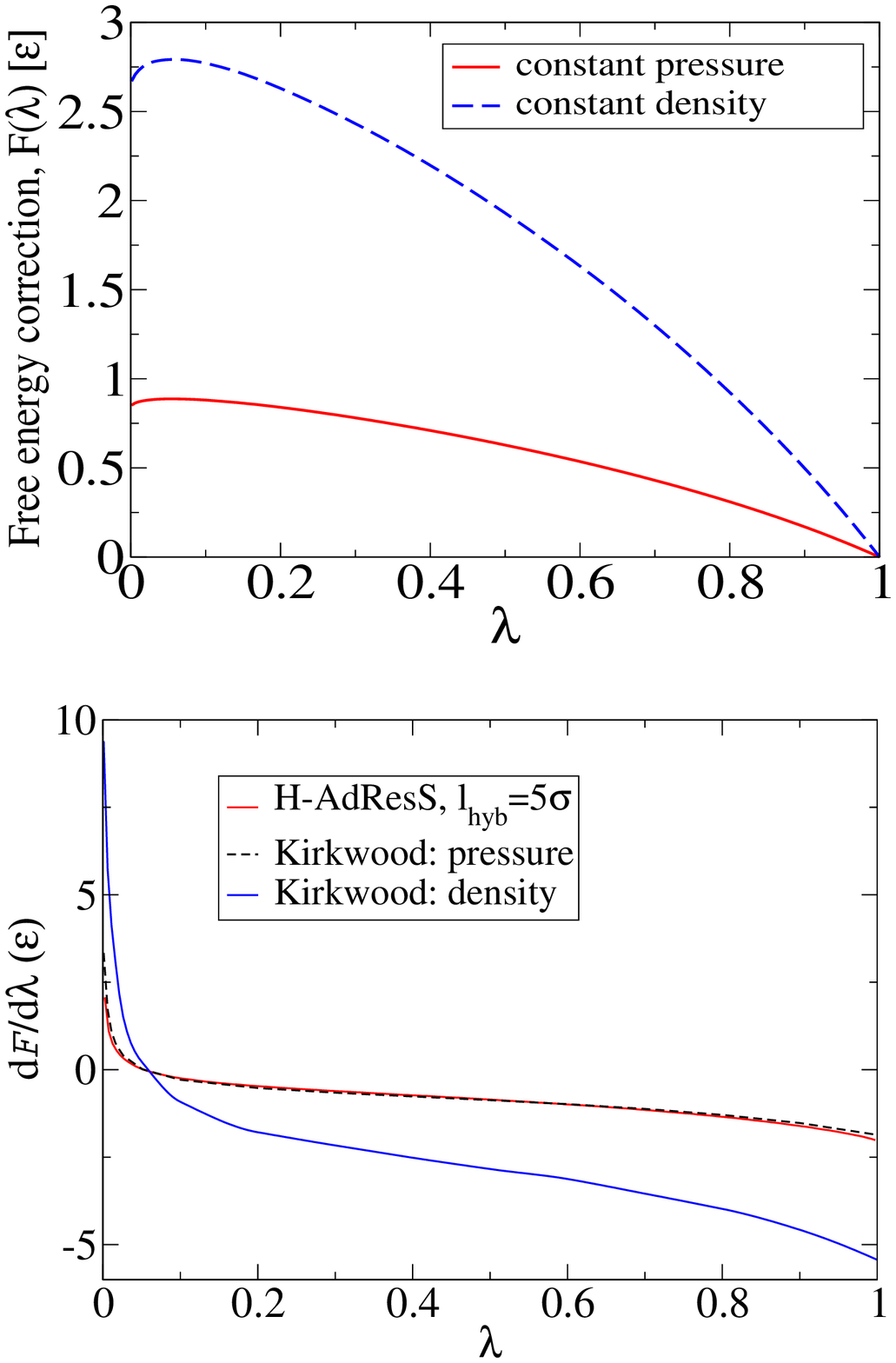}
  \caption{(Top) The  FEC $\calf$  evaluated from  Kirkwood TI  for constant
    pressure  and  constant  density. (Bottom) Derivatives  of Kirkwood free
    energies. In the constant pressure case, the \hadr  FEC derivative
    $\calf'$ is compared with Kirkwood's result. In  this case,
    the   total  Helmholtz  free   energy  difference   (Kirkwood)  is
    $\calf^K(0)=0.85(2)$   while    the   iterative   \hadr   provides
    $\calf(0)=0.86(4)$.}
\label{FEC_NF}
\end{figure}

\begin{widetext}
\begin{figure}[t]
  \includegraphics[scale=0.6]{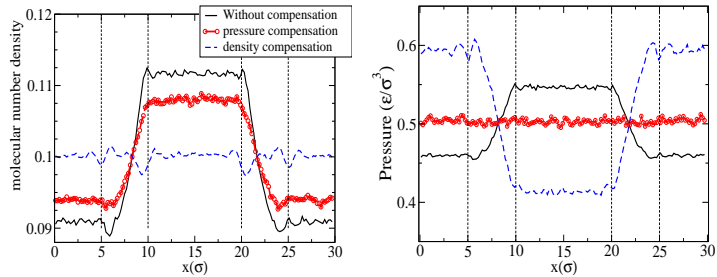}
  \caption{The  density  (left) and  pressure  (right) obtained  using
    \hadr in  simulations of tetrahedral molecules  with the non-fitted
    CG  potential.    The  corresponding  FEC  $\calf$   is  shown  in
    Fig. \ref{FEC_NF}}
\label{fig:nfp}
\end{figure}
%\newpage
\end{widetext}

We  start by  presenting  the free  energy  differences, pressure  and
density  profiles obtained  for the  three cases  considered (constant
pressure and constant  density FEC and no FEC)  of a tetrahedral fluid
facing  a  non-fitted CG  fluid.   These  results  are shown  in  Fig.
\ref{FEC_NF}   (FEC)    and   Fig.    \ref{fig:nfp}    (pressure   and
densities). Note that in this  case the Kirkwood free energy $\calf^K$
is  practically  equal  to   the  constant  pressure  FEC  correction,
reflecting  again  the  strong   connection  of  \hadr  with  standard
statistical  mechanics.   We  will  in  fact hereafter  focus  on  the
constant  density  FEC  and  on its  iterative  evaluation.   Constant
density results of Fig.  \ref{FEC_NF} and \ref{fig:nfp}, obtained with
the Kirkwood  route $\calf^K$, reveal  a relatively large  free energy
difference between  both fluids, of about  $\calf(0) \simeq 2.7\,k_BT$
{\em  per  molecule}.   Under   no-FEC  contribution,  this  leads  to
substantial deviations  in density and pressure  across the simulation
box as reflected in Fig. \ref{fig:nfp}.

\begin{figure}[ht!]
  \includegraphics[scale=0.7]{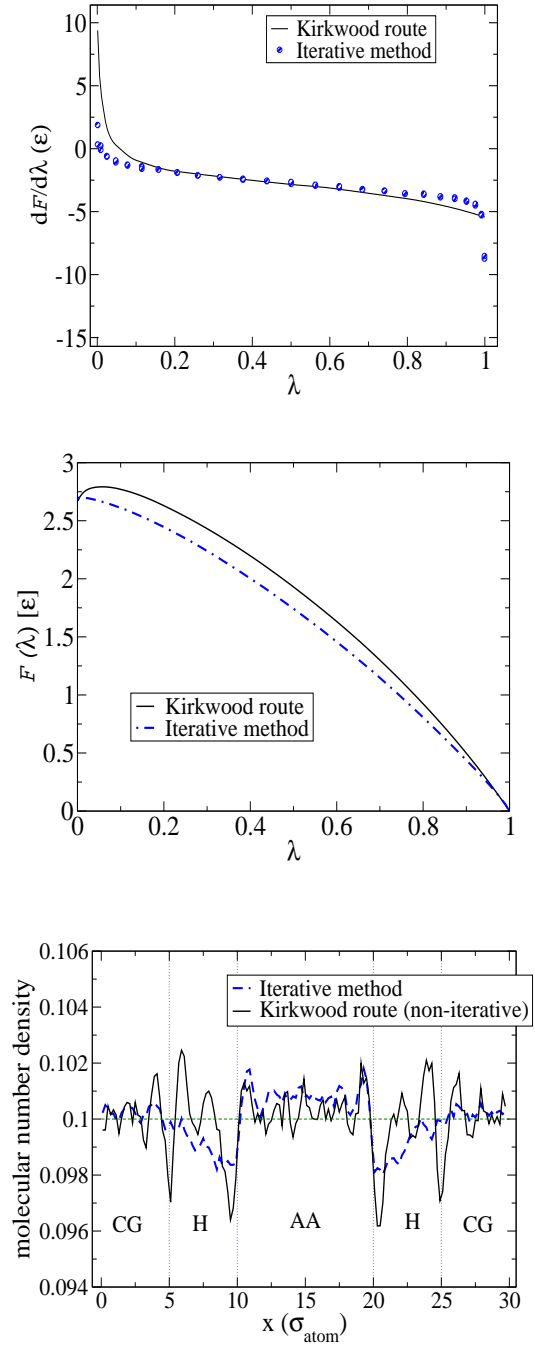}
  \caption{Free  energies  (top)  and  density profiles  (bottom)  for
    constant    density   \hadr    simulations   of    non-fitted   CG
    potentials.  The   Kirkwood  free  energy   $\calf^K(\lambda)$  is
    compared  with the relaxation  algorithm of  Eqs.
(\ref{F*})-(\ref{Fcorrect})  for   the   FEC.    Kirkwood  total   free   energy   jump   is
    $\calf^{K}(0)=2.67(0)$  and  compares quite  well  with the  \hadr
    iterative result $\calf(0)=2.69(7)$. }
\label{fig:nfd}
\end{figure}

\begin{figure}[hb!]
  \includegraphics[scale=0.3]{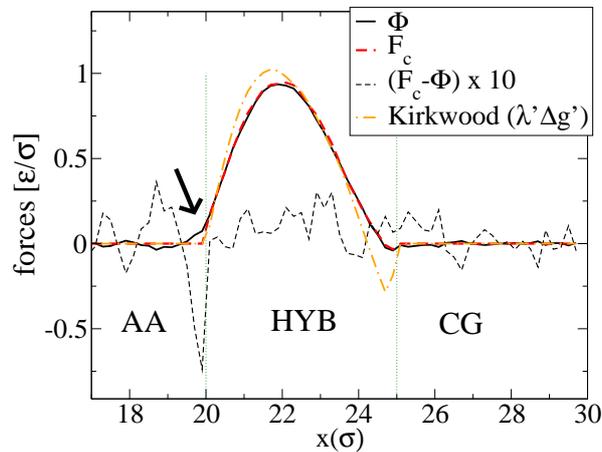}
  \caption{Details of  the force balance of Eq.  (\ref{1Dnconst}) at one
    of the hybrid layers of  a constant density \hadr simulation. The
    iterative evaluation of $\calf'$  was performed with the algorithm
    of Eqs. (\ref{F*})-(\ref{Fcorrect}) 
 using $\Delta t=0.005$, $N_f=5000$ and $\hat{\alpha}=0.01$
and $\Phi=\lambda^{\prime}(x) \left(u^1(x)-u^0(x)\right)/n(x) + g(x)/n(x)$ [see Eq. (\ref{1Dnconst})].}
\label{fig:nfd_fb}
\end{figure}

\subsubsection{Iterative constant density FEC}

  Fig.  \ref{fig:nfd}  compares the  results for $\calf'$  and $\calf$
  using the  iterative evaluation in  Eqs. (\ref{F*})-(\ref{Fcorrect})
  and the  Kirkwood TI in  Eq. (\ref{calFcdens}) for  constant density
  field.   The  first  thing  to highlight  from  Fig.   \ref{fig:nfd}
  (top,half)  is  that although  $\calf'(\lambda)$  (and its  integral
  $\calf(\lambda)$) differ substantially, the {\em overall} \hadr free
  energy difference $\calf(0)$ results to  be equal to the Kirkwood TI
  value.   For the  reasons  explained before,  this  is an  important
  result.  Second,  the density  profile resulting from  the iterative
  protocol  are   not  completely  flat,   although  the  oscillations
  deviating from  the mean density  are softer and smaller  than those
  obtained  from Kirkwood  $\calf^K$ (maximum  density  deviations are
  $2\%$ while about $5\%$ for  Kirkwood).  To understand the origin of
  the  density  differences  resulting  from  the  iterative  protocol
  (\ref{F*})-(\ref{Fcorrect})  we plot  in  Fig. \ref{fig:nfd_fb}  the
  terms  involved in the  force balance  over the  $x$-direction.  The
  system's  average  force  per  molecule  [RHS of Eq.
  \ref{1Dnconst})]  is  compared with  the  imposed compensation  force
  $\ffec$.  Density  variations along  $x$ arise with  any difference
  between  both terms; from  Eq.  (\ref{gibbs2}),  such difference  is
  precisely $k_BT d\ln n/dx$ and for clarity it has been amplified by
  a factor  10 in  Fig.  \ref{fig:nfd_fb}.  Indeed,  $\ffec=0$ inside
  the atomistic  domain but due to  the small width  of the transition
  layer and the  sharp decay to zero of  $\ffec(x)$ (particularly near
  $\lambda=1$, indicated with an arrow in Fig.  \ref{fig:nfd_fb}) the fluid is compressed and
  creates density  oscillations. It seems reasonable  that the density
  oscillations are  larger where the difference  in compressibility is
  (i.e.     near   the    atomistic   border,    $\lambda=1$).    Fig.
  \ref{fig:nfd_fb}  shows that the  transition of  $\ffec$ to  zero is
  softer  at $\lambda=0$, where  the density  profile is  also softer.
  These  observations   indicate  two  things:   first,  that  density
  variations  should  eventually decrease  with  increasing $\lh$  (by
  allowing  smaller  values  of  $|d\ffec/dx|$ within  the  transition
  layer)  and second, that  there might  also be  an optimal  shape of
  $\lambda(x)$. A  study of  these issues is  however left  for future
  work.

\section{Conclusions}
\label{sec:conclu}
This  work presented the statistical  mechanics foundations  of \hadr 
\cite{Potestio:PRL:110,Potestio:PRL:111}.
Because the method is based  on a Hamiltonian, the standard techniques
of Statistical Mechanics  allow one to obtain a  wealth of information
about  the thermodynamics  of AA  and CG  models.  The  Hamiltonian in
\hadr is an  interpolation of the actual microscopic  potential with a
CG representation of the system in terms of blobs. In this way, when a
blob moves from the AA region to the CG region its interactions change
accordingly.   We have  shown  why and  how  \hadr can  be adapted  to
``connect''  two different fluid  models (here  the atomistic  and the
coarse-grained models)  by keeping both  to coexist in the  same fixed
ensemble (for instance,  same density or same pressure)  over the same
simulation box.   The work required to  do that is  precisely the free
energy compensation $\calf$ which  is the central ingredient of \hadr.
We  have proved  that  $\calf(\lambda)$ is close to the  free
energy  difference obtained  from  Kirkwood thermodynamic  integration
 $\calf^K(\lambda)$ and that both  energies are equal in the limit
of   local  thermodynamic  equilibrium   (in  practice,  wide  enough
transition layers).  We have developed schemes to iteratively evaluate
the free energy correction  under either constant pressure or constant
density simulations.  This iterative  route has several benefits.  The
first  is a  practical  one, because  it  avoids the  extra burden  of
implementing Kirkwood thermodynamic integration  each time a FEC needs
to be evaluated. Moreover, iterative evaluation of $\calf$ will permit
to  self-adapt the FEC  under a  (slow enough)  thermodynamic process.
It is important to stress that the {\em overall} free energy
jump in  \hadr is a thermodynamic quantity which
does not depend  on the shape or width of the
transition layer. 
This is confirmed by simulation results
which agree withing error bars with the Kirkwood TI 
free energy evaluation and indeed explains
the good performance  of Kirkwood  TI
approximations       to   $\calf$       used       in       Refs.
\cite{Potestio:PRL:110,Potestio:PRL:111}. 
The  limits  and
potentiality  of \hadr  as  a flexible,  fast  and self-adaptive  free
energy  estimator will surely deserve  further studies
on denser and  more disparate systems.
% and further studies of \hadr thermodynamics.
%by itself, become
%another route to free energy difference estimation; flexible, fast and
%self-adaptive  to any  thermodynamic process.   

The  emphasis in the  present paper  has been  on \textit{equilibrium}
Statistical  Mechanics.   In  order  to  look  at  problems  in  which
\textit{dynamics}  is of importance,  it is  necessary to  include the
possibility in  the algorithm of  interpolating the full  CG dynamics.
In addition to  the CG potential of interaction,  the full CG dynamics
requires the  presence of friction  and stochastic forces in  order to
fully  account for  eliminated degrees  of  freedom in  the CG  region
\cite{Hijon2009}.   As it  is well-known,  the  equilibrium properties
should not be affected by the presence of these additional forces that
are, however,  crucial in non-equilibrium or  dynamic situations. This
further development is left for future work.

\section{Acknowledgments}
We acknowledge the KAVLI Institute in Santa Barbara where this work
was  initiated  for  its  hospitality and  support.   MINECO  provided
support      through       projects      FIS2010-22047-C05-01      and
FIS2010-22047-C05-03.   Comunidad Aut\'onoma  de Madrid  has financially
supported this work through the project MODELICO.

\begin{widetext}
\section{Appendix: Local equilibrium in the transition layer}
\label{localeq}
The exact  result (\ref{nablaAr}) leads to an  interesting result when
the typical length  of variation of $\lambda({\bf r})$  is much larger
than the  typical length of decay  of the correlations.  In this case,
because in the length scale in which the correlation decays, the field
$\lambda({\bf r}')$ hardly changes, we may approximate (\ref{nablaAr})
by  taking  $\nabla  \lambda({\bf  r}')\approx\nabla\lambda({\bf  r})$
outside the integral as follows
\begin{align}
  \nabla  \left\langle  A_{{\bf r}}\right\rangle^{[\lambda]}  &\approx-\beta\nabla\lambda({\bf  r})
  \int  d{\bf r}'\left\langle  \delta A_{\bf
      r}(u^1_{{\bf    r}'}-u^0_{{\bf    r}'}+{\cal    F}'(\lambda({\bf
      r}'))n_{{\bf   r}'})  \right\rangle^{[\lambda]}  \label{nablaArloc}
\end{align} 
This approximation is equivalent to
set, in Eq. (\ref{na})
\begin{align}
\nabla \left\langle A_{{\bf r}}\right\rangle^{[\lambda]}
&\approx \nabla\lambda({\bf r}) \int d{\bf r}'\frac{\delta }{{\delta \lambda({\bf
r}')}}\left\langle A_{{\bf r}}\right\rangle^{[\lambda]}
\label{na2}
\end{align}
Now, let us  consider the average of the  local function $\left\langle
  A_{{\bf   r}}\right\rangle^{[\lambda]}$,  when   $\lambda({\bf  r})$
changes smoothly. Consider the following rewriting of the Hamiltonian
\begin{align}
  H_{[\lambda]}(r,p)&=H_{\lambda({\bf r})}+\delta H_{[\lambda]}
\end{align}
where we have added and subtracted a $\lambda({\bf r})$ term by defining
\begin{align}
H_{\lambda({\bf r})}&\equiv\sum_i\frac{{\bf p}_i^2}{2m_i}
+\sum_\mu^M V^{\rm intra}_\mu(r)+
\lambda({\bf r})\sum_{\mu}^MV^{1}_{\mu}(r)
+(1-\lambda({\bf r}))\sum_{\mu}^M V^{0}_{\mu}(R)
+\sum_\mu^M\calf (\lambda({\bf r}))
\nonumber\\
\delta H_{[\lambda]}&\equiv
\sum_{\mu}^M(\lambda(\hat{\bf R}_\mu)-\lambda({\bf r}))V^{1}_{\mu}(r)
-\sum_{\mu}^M(\lambda(\hat{\bf R}_\mu)-\lambda({\bf r}))V^{0}_{\mu}(R)
+\sum_\mu^M\calf(\lambda(\hat{\bf R}_\mu))-\sum_\mu^M\cal F(\lambda({\bf r}))
\nonumber\\
&=
\int d{\bf r}'(\lambda({\bf r}')-\lambda({\bf r}))u^{1}_{{\bf r}'}(r)
-\int d{\bf r}'(\lambda({\bf r}')-\lambda({\bf r}))u^{0}_{{\bf r}'}(R)
+\int d{\bf r}'(\cal F(\lambda({\bf r}'))-\cal F(\lambda({\bf r})))
\hat{n}_{{\bf r}'}(r)
\label{69}
\end{align}
Clearly, $H_{\lambda({\bf r})}$ is the Hamiltonian of a constant switching field where the value
of the constant is picked to be the local value $\lambda({\bf r})$. 
We can now consider the average of a local function of the form
\begin{align}
\langle \hat{A}_{\bf r}\rangle^{[\lambda]}&=
\int dz\frac{1}{Z[\lambda]}\exp\{-\beta H_{[\lambda]}\}\sum_\mu^M A_\mu\delta({\bf r}-{\bf R}_\mu)
\end{align}
By expanding the exponential with respect to $\delta H_{[\lambda]}$  we have
\begin{align}
\langle \hat{A}_{\bf r}\rangle^{[\lambda]}&=\langle \hat{A}_{\bf r}\rangle^{\lambda=\lambda({\bf r})}
+\langle \delta H_{[\lambda]}\delta\hat{A}_{\bf r}\rangle^{\lambda=\lambda({\bf r})}+\cdots
\end{align}
By using the definition (\ref{69}), we have
\begin{align}
  \langle \delta H_{[\lambda]}\delta\hat{A}_{\bf r}\rangle^{\lambda=\lambda({\bf r})}&=
\int d{\bf r}'(\lambda({\bf r}')-\lambda({\bf r}))  \langle u^{1}_{{\bf r}'}(r)\delta\hat{A}_{\bf r}\rangle^{\lambda=\lambda({\bf r})}
-\int d{\bf r}'(\lambda({\bf r}')-\lambda({\bf r}))\langle u^{0}_{{\bf r}'}(R)\delta\hat{A}_{\bf r}\rangle^{\lambda=\lambda({\bf r})}
\nonumber\\
&+\int d{\bf r}'(\calf(\lambda({\bf r}'))-\calf(\lambda({\bf r})))
\langle\hat{n}_{{\bf r}'}(r)\delta\hat{A}_{\bf r}\rangle^{\lambda=\lambda({\bf r})}
\end{align}
\end{widetext}
It  is apparent that  if the switching field does not changes much on the
length scale of decay of the correlations, all the above contributions may be neglected
and we have 
\begin{align}
\langle    \hat{A}_{\bf    r}\rangle^{[\lambda]}   \approx    \langle
\hat{A}_{\bf r}\rangle^{\lambda=\lambda({\bf r})}  
\label{AAaproxApp}
\end{align}
This is a very natural result that tells that when the switching field
does  not  vary appreciably  in  the  length  scale of  the  molecular
correlations, the average of a  local function in the spatially varying
switching  field is  very well  approximated  with the  average at  a
constant  value of the  switching field  with the  local value  at the
point ${\bf r}$ that we  are considering.  By using this approximation
in Eq. (\ref{na2}), we obtain finally
\begin{align}
\nabla  \langle     \hat{A}_{\bf  r}\rangle^{[\lambda]}       
&\approx    \nabla\lambda({\bf r}) \int d{\bf r}'\frac{\delta }{{\delta \lambda({\bf
r}')}}\left\langle A_{{\bf r}}\right\rangle^{\lambda=\lambda({\bf r})}
\nonumber\\
&= \nabla\lambda({\bf r}) 
\left.\frac{d}{d\lambda}\left\langle A_{{\bf r}}\right\rangle^{\lambda}\right|_{\lambda=\lambda({\bf r})}
\int d{\bf r}'\frac{\delta \lambda({\bf r})}{{\delta \lambda({\bf r}')}}
\nonumber\\
&=
\frac{d}{d\lambda}
\left.   \langle       \hat{A}_{\bf  r}\rangle^{\lambda}\right|_{\lambda=\lambda({\bf   r})}
\nabla \lambda({\bf r})
\label{le1}
\end{align}
This  expression allows one to express  gradients of  local  functions as
simply  proportional  to  the  gradients  of  the  switching  function
whenever the  switching function changes smoothly on  the length scale
of correlations of  the CoM variables.  Eq. (\ref{le1})  could be very
roughly  interpreted   as  a  sort  of  ``chain   rule''  where  space
derivatives are expressed in terms  of derivatives with respect to the
switching field.  The results  (\ref{AAaprox}) and (\ref{le1})  will be
referred as  the local equilibrium approximation for  the averages and
its gradients.

\begin{widetext}
\section{Appendix: The force ${\bf F}_\mu$}
\label{Ap:force}
In this appendix we compute explicitly the force
\begin{align}
\hat{\bf F}_\mu&
=-\sum_i\delta_\mu(i) \frac{\partial}{\partial {\bf r}_i}
\left[\sum_\nu V^{\rm intra}_\nu(r)+
\sum_{\nu}\lambda(\hat{\bf R}_\nu)V^{1}_{\nu}(r)
+\sum_{\nu}(1-\lambda(\hat{\bf R}_\nu))V^{0}_{\nu}(R)
+\sum_\nu\calf(\lambda(\hat{\bf R}_\nu))\right]
\end{align}
Consider the intra  potential energy of molecule $\nu$ which is defined as
\begin{align}
  V^{\rm intra}_\nu(r) &=\frac{1}{2}\sum_{i'j'}\delta_\nu(i')\delta_\nu(j')
\phi^{\rm intra}(r_{i'j'})
\end{align}
where $\phi^{\rm intra}(r_{i'j'})$ is  the pair potential of particles
$i',j'$ due to intramolecular interactions. Then
\begin{align}
  -\sum_i\delta_\mu(i) \frac{\partial}{\partial {\bf r}_i}
\sum_\nu V^{\rm intra}_\nu(r)&=
  -\sum_i\delta_\mu(i) \frac{\partial}{\partial {\bf r}_i}\sum_\nu
\frac{1}{2}\sum_{i'j'}\delta_\nu(i')\delta_\nu(j')
\phi^{\rm intra}(r_{i'j'})
\nonumber\\
&=
  -\sum_i\delta_\mu(i) \sum_\nu
\frac{1}{2}\sum_{i'j'}\delta_\nu(i')\delta_\nu(j')
\frac{\partial}{\partial {\bf r}_i}\phi^{\rm intra}(r_{i'j'})
\nonumber\\
&=
\sum_i\delta_\mu(i) \sum_\nu
\frac{1}{2}\sum_{i'j'}\delta_\nu(i')\delta_\nu(j')
f^{\rm intra}(r_{i'j'}){\bf e}_{i'j'}(\delta_{ii'}-\delta_{ij'})
\nonumber\\
&=
\sum_i\delta_\mu(i) \sum_\nu
\sum_{i'j'}\delta_\nu(i')\delta_\nu(j')
f^{\rm intra}(r_{i'j'}){\bf e}_{i'j'}\delta_{ii'}
\nonumber\\
&=
\sum_i\delta_\mu(i) \sum_\nu
\sum_{j'}\delta_\nu(i)\delta_\nu(j')
f^{\rm intra}(r_{ij'}){\bf e}_{ij'}
\nonumber\\
&=
\sum_i \sum_\nu
\sum_{j'}\delta_{\mu\nu}\delta_\nu(i)\delta_\nu(j')
f^{\rm intra}(r_{ij'}){\bf e}_{ij'}
\nonumber\\
&=
\sum_{ij'}\delta_\mu(i)\delta_\mu(j')
f^{\rm intra}(r_{ij'}){\bf e}_{ij'}=0
\end{align}
because ${\bf e}_{ij}=-{\bf e}_{ji}$ and the indices are dummy.
Indeed the total force on the molecule due to internal forces vanishes.
Consider now the term
\begin{align}
-\sum_i\delta_\mu(i) \frac{\partial}{\partial {\bf r}_i}\lambda(\hat{\bf R}_\nu)  
&=-\sum_i\delta_\mu(i)\nabla \lambda(\hat{\bf R}_\nu)  \frac{\partial}{\partial {\bf r}_i}
\sum_{i'}\delta_\nu(i')\frac{m_{i'}}{m_\nu}{\bf r}_{i'}
\nonumber\\
&=-\sum_i\delta_\mu(i)\nabla \lambda(\hat{\bf R}_\nu) 
\sum_{i'}\delta_\nu(i')\frac{m_{i'}}{m_\nu}\delta_{ii'}
=-\nabla \lambda(\hat{\bf R}_\nu) 
\delta_{\mu\nu}
\end{align}
Next, the term
\begin{align}
  -\sum_i\delta_\mu(i) \frac{\partial}{\partial {\bf r}_i}
V^{1}_{\nu}(r)
&=-\sum_i\delta_\mu(i) \frac{\partial}{\partial {\bf r}_i}
\frac{1}{2}\sum_{\nu'\neq\nu}\sum_{i'j'}\delta_{\nu}(i')\delta_{\nu'}(j')\phi^{\rm inter}(r_{i'j'})
\nonumber\\
&= 
\frac{1}{2}\sum_{\nu'\neq\nu}\sum_{i i'j'}\delta_\mu(i)\delta_{\nu}(i')\delta_{\nu'}(j')
{\bf F}^{1}_{i'j'}(\delta_{ii'}-\delta_{ij'})
\end{align}
where we have introduced the force ${\bf F}^{1}_{i'j'}$ that atom $j'$ exerts on atom $i'$. Therefore
\begin{align}
\nonumber\\
  -\sum_i\delta_\mu(i) \frac{\partial}{\partial {\bf r}_i}
V^{\rm inter}_{\nu}(r)
&= 
\frac{1}{2}\sum_{\nu'\neq\nu}\sum_{i j'}\delta_\mu(i)\delta_{\nu}(i)\delta_{\nu'}(j')
{\bf F}^{1}_{ij'}
-
\frac{1}{2}\sum_{\nu'\neq\nu}\sum_{i i'}\delta_\mu(i)\delta_{\nu}(i')\delta_{\nu'}(i)
{\bf F}^{1}_{i'i}
\nonumber\\
&= 
\delta_{\mu\nu}\frac{1}{2}\sum_{\nu'\neq\nu}\sum_{i j}\delta_\mu(i)\delta_{\nu'}(j)
{\bf F}^{1}_{ij}
-
\frac{1}{2}\sum_{\nu'\neq\nu}\delta_{\mu\nu'}\sum_{i j}\delta_\mu(i)\delta_{\nu}(j)
{\bf F}^{1}_{ji}
\nonumber\\
&= 
\delta_{\mu\nu}\frac{1}{2}\sum_{\nu'\neq\nu}{\bf F}^{1}_{\mu\nu'}
+
\frac{1}{2}\sum_{\nu'\neq\nu}\delta_{\mu\nu'}{\bf F}^{1}_{\mu\nu}
\nonumber\\
&=\sum_{\nu'\neq\nu}{\bf F}^{1}_{\nu\nu'}(R_{\nu\nu'})
\frac{1}{2}\left[\delta_{\mu\nu}-\delta_{\mu\nu'}\right]
\end{align}
where we have introduced the force that molecule $\mu $ exerts on molecule $\nu$ as
\begin{align}
{\bf F}^{1}_{\mu\nu}&\equiv
\sum_{i j}\delta_\mu(i)\delta_{\nu}(j)
{\bf F}^{1}_{ij}
\end{align}

Next, the term
\begin{align}
  -\sum_i\delta_\mu(i) \frac{\partial}{\partial {\bf r}_i}
V^{0}_{\nu}(R)&=
  -\sum_i\delta_\mu(i) \frac{\partial}{\partial {\bf r}_i}
\frac{1}{2}\sum_{\nu'}V^{0}_{\nu\nu'}(R)
\nonumber\\
&= 
\frac{1}{2}\sum_{\nu'}F^{\rm cm}_{\nu\nu'}(R_{\nu\nu'})
 \sum_i\delta_\mu(i) \frac{\partial{R}_{\nu\nu'}}{\partial {\bf r}_i}
\nonumber\\
&= 
\frac{1}{2}\sum_{\nu'}F^{\rm cm}_{\nu\nu'}(R_{\nu\nu'})
 \sum_i\delta_\mu(i) \frac{\partial{R}_{\nu\nu'}}{\partial {\bf r}_i}
\end{align}
where we assumed pair-wise interactions. Then
\begin{align}
   \sum_i\delta_\mu(i) \frac{\partial{\bf R}_{\nu}}{\partial {\bf r}_i}&=
   \sum_i\delta_\mu(i) \frac{\partial}{\partial {\bf r}_i}
\sum_{i'}\delta_\nu(i')
\frac{m_{i'}}{m_\nu}{\bf r}_{i'}
\nonumber\\
&=
   \sum_i\delta_\mu(i) 
\sum_{i'}\delta_\nu(i')
\frac{m_{i'}}{m_\nu}{\bf 1}\delta_{ii'}
=
   \sum_i\delta_\mu(i) 
\delta_\nu(i)
\frac{m_{i}}{m_\nu}{\bf 1}=\delta_{\mu\nu}{\bf 1}
\end{align}

\begin{align}
   \sum_i\delta_\mu(i) \frac{\partial{R}_{\nu\nu'}}{\partial {\bf r}_i}&=
 {\bf e}_{\nu\nu'}\cdot   \sum_i\delta_\mu(i)\frac{\partial{\bf R}_{\nu\nu'}}{\partial {\bf r}_i}
= {\bf e}_{\nu\nu'}\left[\delta_{\mu\nu}-\delta_{\mu\nu'}\right]
\end{align}
then
\begin{align}
  -\sum_i\delta_\mu(i) \frac{\partial}{\partial {\bf r}_i}
V^{0}_{\nu}(R)
&= 
\frac{1}{2}\sum_{\nu'}F^{\rm cm}_{\nu\nu'}(R_{\nu\nu'})
 \sum_i\delta_\mu(i) \frac{\partial{R}_{\nu\nu'}}{\partial {\bf r}_i}
\nonumber\\
&= 
\frac{1}{2}\sum_{\nu'}F^{\rm cm}_{\nu\nu'}(R_{\nu\nu'})
 {\bf e}_{\nu\nu'}\left[\delta_{\mu\nu}-\delta_{\mu\nu'}\right]
\nonumber\\
&= 
\frac{1}{2}\sum_{\nu'}{\bf F}^{0}_{\nu\nu'}(R_{\nu\nu'})
\left[\delta_{\mu\nu}-\delta_{\mu\nu'}\right]
\end{align}

In summary, we have 
\begin{align}
\hat{\bf F}_\mu&\equiv
-\sum_i\delta_\mu(i) \frac{\partial}{\partial {\bf r}_i}
\left[\sum_\nu V^{\rm intra}_\nu(r)+
\sum_{\nu}\lambda(\hat{\bf R}_\nu)V^{\rm inter}_{\nu}(r)
+\sum_{\nu}(1-\lambda(\hat{\bf R}_\nu))V^{0}_{\nu}(R)
+\sum_\nu\calf(\lambda(\hat{\bf R}_\nu))\right]
\end{align}
and have to substitute in this expression the following results
\begin{align}
  -\sum_i\delta_\mu(i) \frac{\partial}{\partial {\bf r}_i}
\sum_\nu V^{\rm intra}_\nu(r)&=0
\nonumber\\
-\sum_i\delta_\mu(i) \frac{\partial}{\partial {\bf r}_i}\lambda(\hat{\bf R}_\nu)  
&=-\nabla \lambda(\hat{\bf R}_\nu) 
\delta_{\mu\nu}
\nonumber\\
  -\sum_i\delta_\mu(i) \frac{\partial}{\partial {\bf r}_i}
V^{\rm inter}_{\nu}(r)
&=\frac{1}{2}\sum_{\nu'}{\bf F}^{\rm intra}_{\nu\nu'}(R_{\mu\nu'})
\left[\delta_{\mu\nu}-\delta_{\mu\nu'}\right]
\nonumber\\
  -\sum_i\delta_\mu(i) \frac{\partial}{\partial {\bf r}_i}
V^{0}_{\nu}(R)
&=\frac{1}{2}\sum_{\nu'}{\bf F}^{0}_{\nu\nu'}(R_{\nu\nu'})
\left[\delta_{\mu\nu}-\delta_{\mu\nu'}\right]
\end{align}
with the result
\begin{align}
\hat{\bf F}_\mu&=
-\left[\sum_{\nu}\sum_i\delta_\mu(i) \frac{\partial}{\partial {\bf r}_i}
\lambda(\hat{\bf R}_\nu)(V^{\rm inter}_{\nu}(r)
-V^{0}_{\nu}(R)-\calf'(\lambda_\nu(R))
)\right]
\nonumber\\
&-\left[\sum_{\nu}\lambda(\hat{\bf R}_\nu)\sum_i\delta_\mu(i) \frac{\partial}{\partial {\bf r}_i}V^{\rm inter}_{\nu}(r)
+\sum_{\nu}(1-\lambda(\hat{\bf R}_\nu))\sum_i\delta_\mu(i) \frac{\partial}{\partial {\bf r}_i}V^{0}_{\nu}(R)
\right]
\nonumber\\
&=-\nabla\lambda(\hat{\bf R}_\mu)(V^{\rm inter}_{\mu}(r)-V^{0}_{\mu}(R)-\calf'(\lambda_\mu(R)))
\nonumber\\
&+\sum_{\nu}\lambda(\hat{\bf R}_\nu)\frac{1}{2}\sum_{\nu'}{\bf F}^{1}_{\nu\nu'}(R_{\mu\nu'})
\left[\delta_{\mu\nu}-\delta_{\mu\nu'}\right]
+\sum_{\nu}(1-\lambda(\hat{\bf R}_\nu))
\frac{1}{2}\sum_{\nu'}{\bf F}^{0}_{\nu\nu'}(R_{\nu\nu'})
\left[\delta_{\mu\nu}-\delta_{\mu\nu'}\right]
\nonumber\\
&=-\nabla\lambda(\hat{\bf R}_\mu)(V^{\rm inter}_{\mu}(r)-V^{0}_{\mu}(R)-\calf'(\lambda_\mu(R)))
\nonumber\\
&+\sum_{\nu}\frac{\lambda(\hat{\bf R}_\mu)+\lambda(\hat{\bf R}_{\nu})}{2}
{\bf F}^{1}_{\mu\nu}(R_{\mu\nu})
+\sum_{\nu}\left(1-\frac{\lambda(\hat{\bf R}_\mu)+\lambda(\hat{\bf R}_{\nu})}{2}\right)
{\bf F}^{0}_{\nu\nu'}(R_{\mu\nu})
\end{align}
We may introduce the following pair force
\begin{align}
  \hat{\bf G}_{\mu\nu}&=
\left[\frac{\lambda(\hat{\bf R}_\mu)+\lambda(\hat{\bf R}_\nu)}{2}\right]
{\bf F}^{1}_{\mu\nu}(R_{\mu\nu})
+\left[1-\frac{\lambda(\hat{\bf R}_\mu)+\lambda(\hat{\bf R}_\nu)}{2}\right]
{\bf F}^{0}_{\mu\nu}(R_{\mu\nu})
\end{align}
The pair force satisfies Newton's Third Law. With this definition we have
\begin{align}
\hat{\bf F}_\mu&=
-\nabla\lambda(\hat{\bf R}_\mu)(V^{\rm inter}_{\mu}(r)-V^{0}_{\mu}(R)-\calf'(\lambda_\mu(R)))
+\sum_{\nu}  \hat{\bf G}_{\mu\nu}
\end{align}
\end{widetext}

\bibliography{bibliohadres,pep_collection}

\end{document}